\documentclass[review]{elsarticle}
\usepackage{lineno,hyperref,graphicx,color}
\modulolinenumbers[5]
\journal{Nuclear Instruments and Methods}
\newif\ifcomment
\newif\ifextrafigs
\newif\ifdraft
\newcommand{\com}[1]{\relax}

\newcommand{\Fig}[1]{Fig.~\ref{#1}}

\begin{document}
\begin{frontmatter}
\title{A Large-Scale Pad-Sensor Based Prototype of the Silicon Tungsten Electromagnetic Calorimeter for the Forward Direction in ALICE at LHC}
\address[utrecht]{Institute for Gravitational and Subatomic Physics (GRASP), Utrecht University/Nikhef, Netherlands}
\address[tsukuba]{University of Tsukuba, Tsukuba, Japan}
\address[tsukuba_tech]{Tsukuba University of Technology, Tsukuba, Japan}
\address[ornl]{Oak Ridge National Laboratory, Oak Ridge, Tennessee, United States}
\address[hiroshima]{Hiroshima University, Higashi Hiroshima, Japan}
\address[nara]{Nara Women's University, Nara, Japan}
%
\author[utrecht]{R.G.E.~Barthel}
\author[tsukuba]{T.~Chujo\corref{mycorrespondingauthor}}
\cortext[mycorrespondingauthor]{Corresponding author}
\ead{chujo.tatsuya.fw@u.tsukuba.ac.jp}
\author[nara]{T.~Hachiya}
\author[tsukuba]{M.~Hatakeyama}
\author[tsukuba]{Y.~Hoshi}
\author[tsukuba_tech]{M.~Inaba}
\author[tsukuba]{Y.~Kawamura}
\author[tsukuba]{D.~Kawana}
\author[ornl]{C.~Loizides}
\author[tsukuba]{Y.~Miake}
\author[nara]{Y.~Minato}
\author[tsukuba]{K.~Nakagawa}
\author[tsukuba,ornl]{N.~Novitzky}
\author[utrecht]{T.~Peitzmann}
\author[utrecht]{M. Rossewij}
\author[nara]{M.~Shimomura}
\author[hiroshima]{T.~Sugitate}
\author[tsukuba]{T.~Suzuki}
\author[tsukuba]{K.~Tadokoro}
\author[tsukuba]{M.~Takamura}
\author[hiroshima]{S.~Takasu}
\author[utrecht]{A.~van den Brink}
\author[utrecht]{M.~van Leeuwen}
\date{\today}

\begin{abstract}
We constructed a large-scale electromagnetic calorimeter prototype as a part of the Forward Calorimeter upgrade project~(FoCal) for the ALICE experiment at the Large Hadron Collider (LHC). 
The prototype, also known as ``Mini FoCal'', consists of 20 layers of silicon pad sensors and tungsten alloy plates with printed circuit boards and readout electronics. 
The constructed detector was tested at the test beam facility of the Super Proton Synchrotron~(SPS) at CERN. 
We obtain an energy resolution of about 4.3\% for electron beams at both 150 and 250 GeV/$c$, which is consistent with realistic detector response simulations.
Longitudinal profiles of electromagnetic shower were also measured and found to agree with the simulations. 
The same prototype detector was installed in the ALICE experimental area about 7.5~m away from the interaction point. 
It was used to measure inclusive electromagnetic cluster energy distributions and neutral-pion candidate invariant mass distributions for pseudo-rapidity of $\eta$=3.7--4.5 in proton--proton collisions at $\sqrt{s}$ = 13 TeV at LHC. 
The measured distributions in different $\eta$ regions are similar to those obtained from PYTHIA simulations. 
\end{abstract}

\end{frontmatter}
\ifdraft
\linenumbers
\fi

\section{Introduction}
\label{sec:intro}
A new instrument in the forward direction, the Forward Calorimeter (FoCal), was proposed~\cite{ALICE:2020mso} for the ALICE experiment at the Large Hadron Collider (LHC) for Run 4. 
It has unique capabilities to constrain the small-$x$ gluon structure of protons and nuclei via very forward measurements~($3.2< \eta< 5.8$) of direct photons, neutral hadrons, jets and their correlations in proton-proton and proton-lead collisions at LHC, as well as J/$\psi$ production in ultra-peripheral heavy-ion collisions~\cite{ALICE:2023fov}. 
For the electromagnetic part of FoCal~(FoCal-E) a Silicon-Tungsten~(Si-W) calorimeter using both Si-pad and Si-pixel layers to discriminate single photons from pairs of photons originating from $\pi^{0}$ decays is foreseen. 
The hadronic part of FoCal~(FoCal-H) is being designed as a conventional sampling hadron calorimeter mainly for jet measurements and isolation of direct photons. 

Si-W calorimeters for the application in large high-energy physics experiments have been discussed since some time, but no full scale detector has yet been constructed. The first prototype for a potential such detector at ILC, the SiW-ECAL physics prototype, was built by the CALICE collaboration in 2006~\cite{CALICE:2008gxs} and very promising performance of this technology was demonstrated~\cite{CALICE:2008kht}.

Recently, more advanced prototypes of Si-W calorimeters using Si-pads of similar granularity~($1.1\times1.1$~cm$^2$ and $0.5\times0.5$~cm$^2$) have been been built and tested by the CALICE and CMS collaborations, respectively~\cite{Kawagoe:2019dzh,CMSHGCAL:2021nyx}. 
Both these efforts are derived from the early CALICE physics prototype, together with the development of dedicated readout electronics, and they show a consistently good performance. 

In 2015, we built and tested a first prototype of FoCal-E~\cite{Awes:2019vfi}.
This prototype, which consisted of 16 silicon pad and 2 pixel layers, showed a reasonable energy linearity (within 3\%) from 0.5 to 50 GeV for electron beams.
The constant term of the energy resolution was determined to be about 10\% consistent with that obtained from a detector simulation, which included realistic noise and dead space.
The sensitive area of the 2015 prototype was about 8$\times$8~$\rm{cm}^2$ Si-pad sensors, with 16 layers of Si-W. 
The pad layers were read out in 4 groups of 4 layers, in which the signals from 4 successive layers in the longitudinal direction were summed up by custom front-end ASICs.

For the present research, a new prototype, called ``Mini FoCal'',  with three times larger sensitive area, 24$\times$8~$\rm{cm}^2$, and 20 layers of Si-W pad layers has been built. 
The main purposes of this development step are to improve on the performance compared to the previous prototype and to demonstrate the feasibility of such a detector in the ALICE experiment under real-life conditions.
Instead of summing up signals for 4 layers in the longitudinal direction, we adopted a new scheme, in which for every layer each pad sensor is read out individually, using the APV25 hybrid-SRS system. 
The intention was to achieve a better energy resolution by allowing for an individual calibration of each sensor and readout card, and to measure the longitudinal shower profile more precisely. 
In 2018, the new sensors to be used in the prototype were tested at the test beam area of the Proton Synchrotron~(PS).
Measurements with the final prototype were performed at the test beam area of the Super Proton Synchrotron~(SPS), and with pp collisions of the LHC in the ALICE experimental area at CERN. 
In this paper, we report on the construction and the performance of this large-scale prototype for the pad layers of the future ALICE FoCal.

\section{Detector prototype}
\label{sec:Detector}
A large-scale prototype for the pad component of the future ALICE FoCal, called ``Mini FoCal'', with 20 Si-W layers has been developed using 60 n-substrate silicon pad sensors. 
In the following, we describe the silicon pad sensor, tungsten alloy plate, flexible printed circuit (FPC) and the readout system in more detail.

\begin{figure}[t!]
\begin{center}
 \includegraphics[width=70mm,clip]{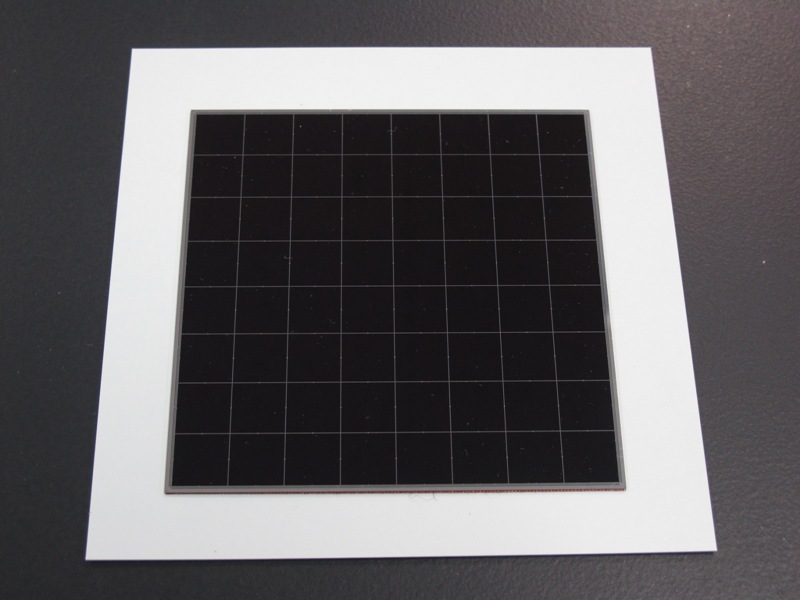}
\end{center}
\caption{The first version of the n-type sensor with $500 {\rm \mu m}$ thickness.}
\label{fig:Sensor500}
\end{figure}
\begin{figure}[t!]
\begin{center}
 \includegraphics[width=120mm,clip]{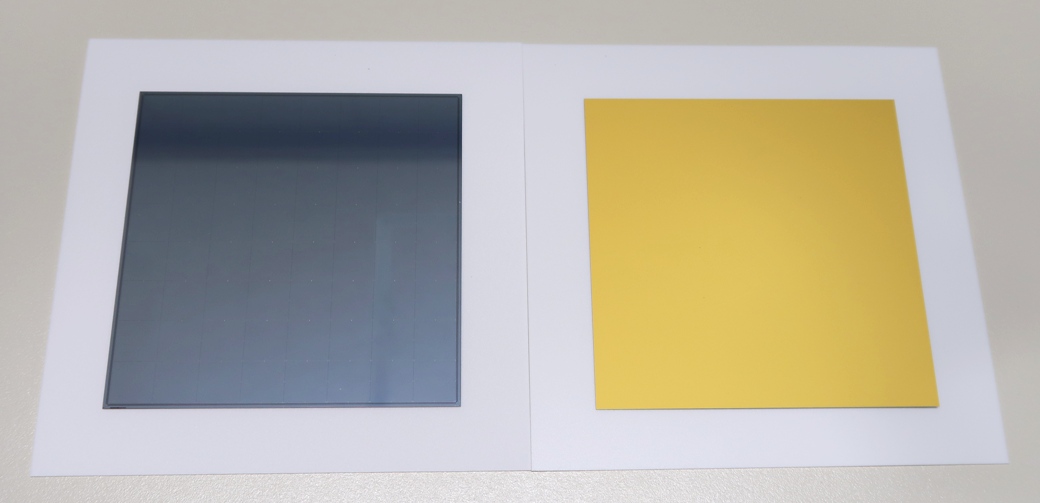}
\end{center}
\caption{The second version of the n-type sensors with $320 {\rm \mu m}$ thickness, which were used for the Mini FoCal. On the left, a sensor is shown with the Al-covered side, on the right a sensor with the opposite, Au-covered one. }
\label{fig:Sensor320}
\end{figure}

\subsection{Silicon pad sensor}
\label{sec:Sensor}

We produced two kinds of custom-made n-substrate silicon PIN photodiode array sensors with the $8\times8$ pad layout, manufactured by Hamamatsu Photonics K.K..  
The first sensors had a thickness of 500~${\rm \mu m}$ and 64 pads without Al cover on the surface as shown in \Fig{fig:Sensor500}. 
The pad pitch was 11.3 ${\rm mm}$, giving a total sensor size of 92.6${\rm mm \times 92.6 mm}$, including the guard rings.  
After several tests on the test bench and in the ELPH test beam line at the Tohoku University in Japan, we decided to procure sensors with improved radiation tolerance and light-shielding properties. 

For improved radiation tolerance, the thickness of the sensor was reduced to 320 ${\rm \mu m}$ including 20 ${\rm \mu m}$ of a non-active area on the back side. 
For better light shielding and also for reduced noise, the surface of the sensor was covered with an aluminium layer as shown in \Fig{fig:Sensor320}.  
The back side of the sensor is coated with a thin gold layer to improve the grounding. 
The leakage current of the sensor was low,  between 1 and 2~${\rm nA}$ per pad at a bias voltage of 100 Volt for most of the pads. 
By design, the full depletion was reached at 35~V and the maximum operating voltage was 220~V. 

\subsection{Absorber}
\label{sec:Tungsten}

\begin{figure}[t!]
\begin{center}
 \includegraphics[width=85mm,clip]{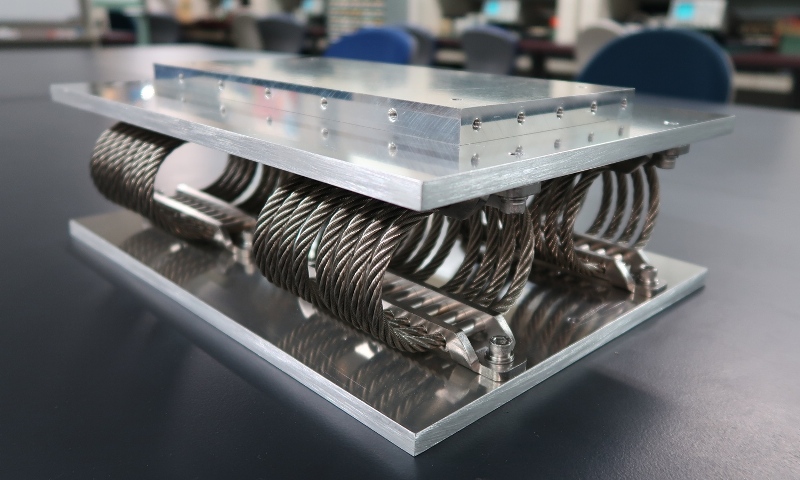}
\end{center}
\caption{A special transportation instrument with shock absorbers for Mini FoCal.}
\label{fig:Transport}
\end{figure}

As absorbers of the Mini FoCal, custom-made non-magnetic heavy-metal plates consisting of a W-alloy with a composition of 94\% W, 4\% Ni and 2\% Cu was adopted.  
In total 20 plates were produced, each with a thickness of 3.5~${\rm mm}$ corresponding to 1 radiation length ${X_0}$.  
Each Tungsten alloy plate measures ${281.0~\rm  mm\times94.0~mm\times3.5~mm}$ and has a weight of 1.65~${\rm kg}$.
Three silicon pad sensors were placed side by side with a gap of $0.2$~mm on the tungsten alloy plate using an electrically conductive adhesive which provides a ground connection for the sensor bias. 
The plates were designed to be slightly larger than the total size of the three mounted sensors in order to prevent mechanical damage of the sensors.  
The flatness of the plates was well controlled in production.  
Since the total weight of 20 plates amounted to about 30~${\rm kg}$, we also developed a heavy-duty transportation instrument with special shock absorbers for shipping from Japan to CERN, Switzerland as shown in \Fig{fig:Transport}. 

\begin{figure}
\centering
\begin{minipage}[t]{0.3\textwidth}
 \centering
 \includegraphics[width=\textwidth,clip]{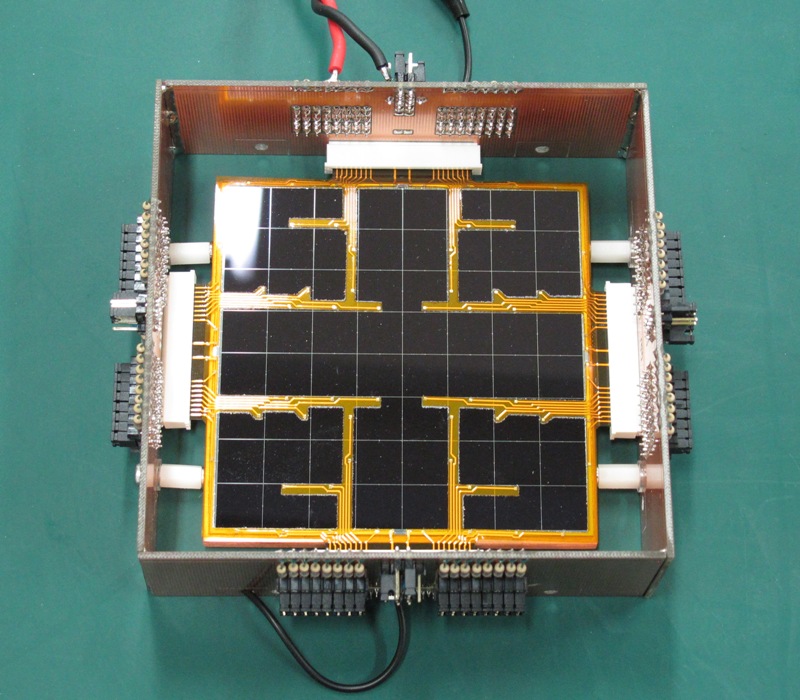}
 \caption{A test setup of the silicon pad sensor using a two-layer FPC.}
 \label{fig:Sensor500test1}
\end{minipage}
\hspace{0.1cm}
\begin{minipage}[t]{0.3\textwidth}
\centering
 \includegraphics[width=\textwidth,clip]{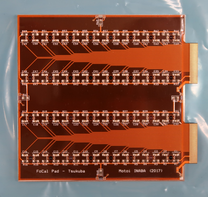}
 \caption{A FPC with bias resistors, high-voltage capacitors and two connectors.}
 \label{fig:Sensor500test2}
\end{minipage}
\hspace{0.1cm}
\begin{minipage}[t]{0.3\textwidth}
\centering
 \includegraphics[width=\textwidth,clip]{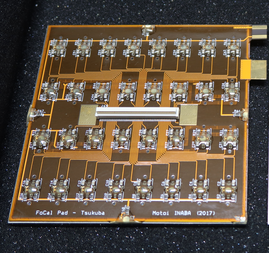}
 \caption{A FPC with bias resistors, high-voltage capacitors and a 130-pin connector.} 
 \label{fig:Sensor320FPC}
\end{minipage}
\end{figure}

\begin{figure}[t!]
\begin{center}
 \includegraphics[width=100mm,clip,trim=10 25 8 25]{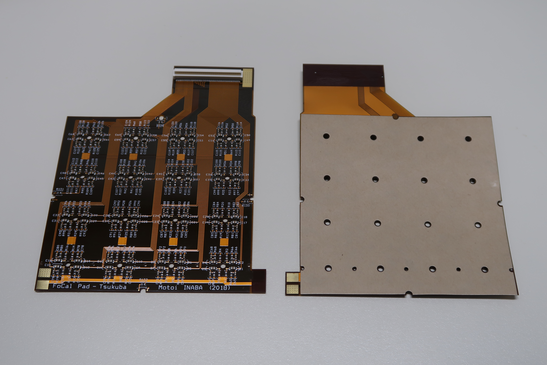}
\end{center}
\caption{Top and bottom of the FPC with the 130-pin connector to one side.}
\label{fig:FPC}
\end{figure}

\subsection{FPC}
\label{sec:FPC}

A flexible printed circuit~(FPC) was used to realize a thin and low-mass signal path from the pads of the sensor to the front-end electronics.  
As a first step, we designed a two-layer FPC to make a connection between the sensor and the outside connection on the test bench as shown in \Fig{fig:Sensor500test1}.  
The FPC and sensor were connected with each other by wire bonding.  
As the second step, we made new versions of the FPCs with a) resistors for the bias voltage, b) high-voltage capacitors for cutting the direct current in the signal path and c) a low-voltage connector mounted on the surface as shown in \Fig{fig:Sensor500test2} and \Fig{fig:Sensor320FPC}. 
The FPC has some holes for wiring, and it provide a bias voltage up to 220 V to 64 pads and the guard ring of the sensor.  

\begin{figure}[t!]
\begin{center}
 \includegraphics[width=85mm,clip]{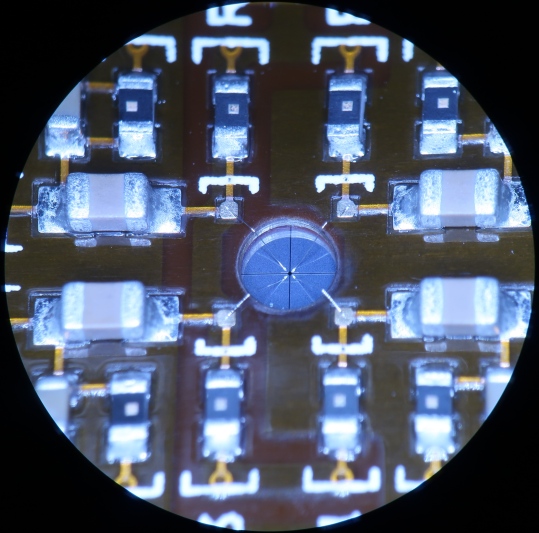}
\end{center}
\caption{Bonding wires from the corner of each of the four silicon pad sensors underneath the FPC routed through the hole in the center of the FPC.}
\label{fig:WireBonding}
\end{figure}

\begin{figure}[t!]
\begin{center}
 \includegraphics[width=85mm,clip]{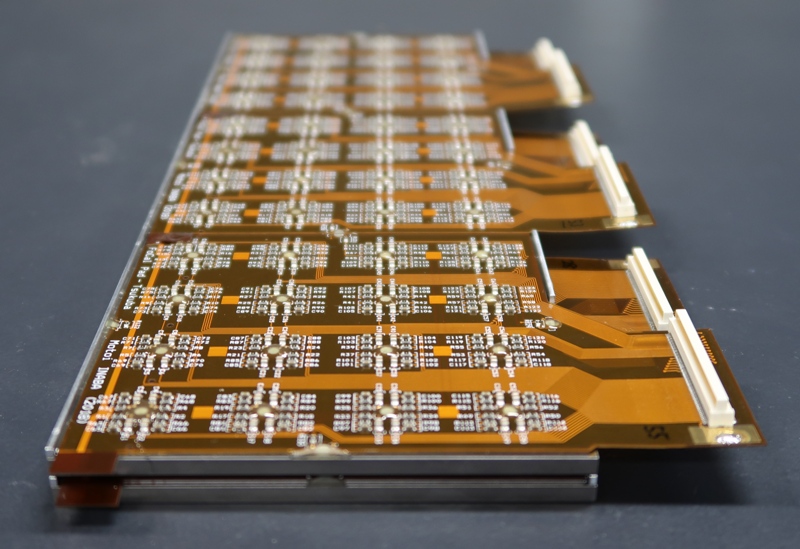}
\end{center}
\caption{A pair of layers for the Mini FoCal.}
\label{fig:LayerPair}
\end{figure}

Since it was important to keep the thickness of each layer to a minimum, we developed a final version of the two-layer FPC which had a 130-pin connector located outside of the sensitive area. 
We made left- and right-handed versions which were alternated in consecutive layers in order to have more space for the connection to the cables.
The FPC was glued on the surface of the silicon pad sensor using an electrically insulating adhesive sheet with some holes as shown in \Fig{fig:FPC}.
All pads and the guard ring of the sensor are electrically connected with the FPC using aluminum bonding wires through holes on the FPC as shown in \Fig{fig:WireBonding}. 
Three FPCs with the silicon pad sensors placed side-by-side on the tungsten alloy plate were connected with each other in order to provide the bias voltage to all FPCs.  
Finally, consecutive Si-W layers make up a pair as shown in \Fig{fig:LayerPair}.
The resulting thickness of a pair layer is $10.3 {\rm mm}$ including two Tungsten alloy plates.

In order to obtain a wider dynamic range, it was decided to attenuate the signal of some layers using additional capacitors on the FPC.  
The 1st and 2nd layer were set to have 1/18 attenuation, while all other layers have 1/180 as attenuation factor. 
This scheme allowed us to cover the required dynamic range with the APV25 readout modules for the 100--250 GeV electron beams.

\subsection{Readout system}
\label{sec:Readout}

\begin{figure}[t!]
\begin{center}
 \includegraphics[width=120mm,clip]{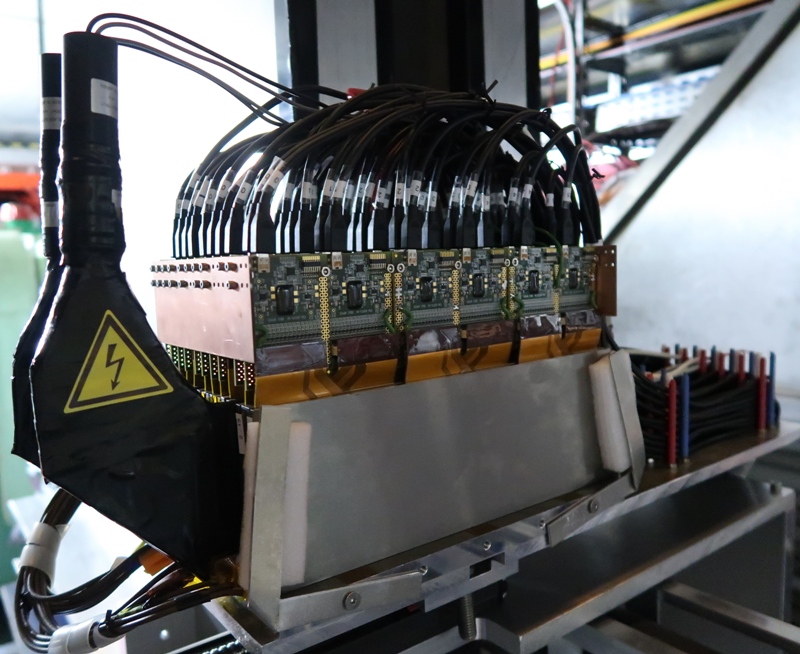}
\end{center}
\caption{The Mini FoCal interfaced with the APV25 hybrid boards.}
\label{fig:MiniFoCal}
\end{figure}

The RD51 APV25 hybrid board was used as front-end electronics with the RD51 Scalable Readout System (SRS) 12-bit ADC-C and FEC v6 cards~\cite{Martoiu:2013aca}.
The APV25 has 128 input channels featuring discharge protection, charge-sensitive amplifier, shaper amplifier and a 192-cell analog  memory pipeline~\cite{French:2001xb}, and it takes samples of analog signals from 64 pads of the sensor through a 130-pin connector at 25 $ \rm ns$ intervals.  
When a trigger signal is identified, 13 analog samples are sent to the ADC card through an HDMI cable and digitized by the ADS5281 chip.  
A waveform of the signal can be reconstructed using the 13~(or more up to 30 programmable) data samples mentioned above and the charge information is calculated.  
A total of 60 APV25 hybrid boards were placed on top of the Si-W layers to read out the signals of 3840 pads in Mini FoCal as shown in \Fig{fig:MiniFoCal}. 

\vspace{5mm}

\section{Simulation setup}
\label{sec:simulation}

The simulation of the detector was done using the Geant4.10.7~\cite{GEANT4:2002zbu} software package. 
We implemented in detail the geometry and the material of the various components of the prototype detector. 
Each layer was described using a $3.5~{\rm mm}$ thick tungsten-alloy plate consisting of 94\% W, 4\% Ni and 2\% Cu. 
The active volume of the sensor was given with a thickness of $320~{\rm \mu m}$.
The glue between W and sensor as well as between sensor and FCP was simulated by a $1.38~{\rm g/cm^{3}}$ density epoxy glue material with a thickness of $110~{\rm \mu m}$ and $130~{\rm \mu m}$, respectively.
A FPC was approximated by a $100~{\rm \mu m}$ thick uniform Cu layer to account for the electronic components. 
In addition, we also included a $1.2~{\rm mm}$ air gap between the layers.

The individual sensors were described as $8\times8$ pads with individual pad sizes of $11.3\times 11.3 ~{\rm mm^{2}}$. 
Each sensor was surrounded by a guard ring which is a passive silicon region around the full circumference of the main sensor with a width of $1.3~{\rm mm}$. 
We also accounted for the $0.2 {\rm mm}$ space between the sensors which was implemented to allow for some tolerance for the placement of the sensors on the W layer. 

\begin{figure}[t!]
\begin{center}
 \includegraphics[width=80mm,clip]{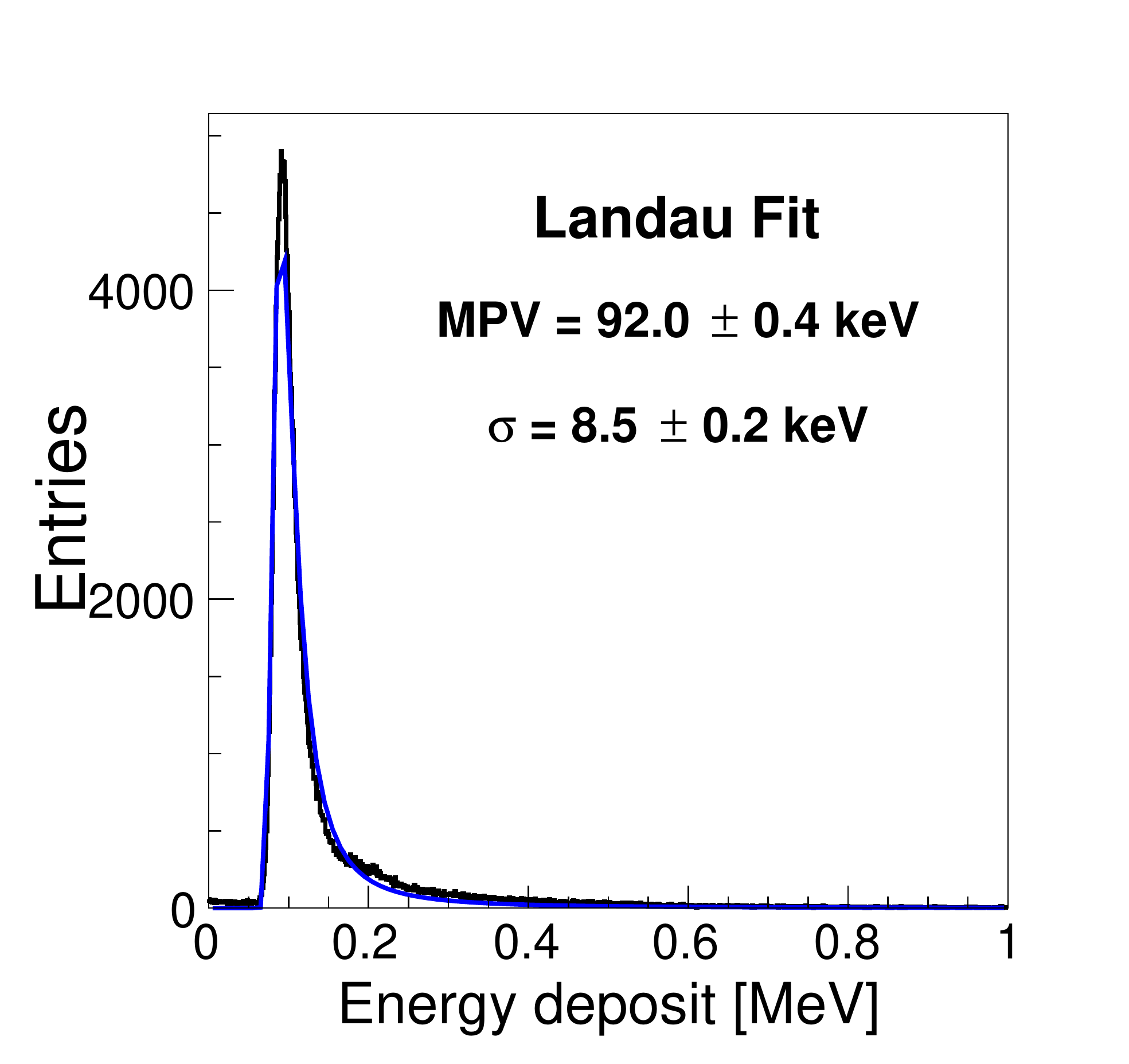}
\end{center}
\vspace{-5mm}
\caption{Simulated MIP response for one layer of the Mini FoCal prototype detector.}
\label{fig:muon}
\end{figure}

The Geant4 simulation team recommends $\rm FTFP\_BERT$ physics list for high energy calorimetry. 
We also tested the other recommended list $\rm QGSP\_BERT$ and found only negligible difference in the detector response. 
To compare with the data, we reconstruct the total deposited energy in each pad of the sensor in the simulations.
For reference to the minimum ionizing particle~(MIP) response of the detector, we simulated muons with 30 GeV energy and reconstructed the response in each pad, as seen in \Fig{fig:muon}. 
In the simulation, a MIP deposits per layer an energy with a most probable value of about 92~keV and a width of 8.5~keV obtained from a fit of a Landau function. 

\begin{figure}[t!]
\begin{center}
 \includegraphics[width=100mm,clip]{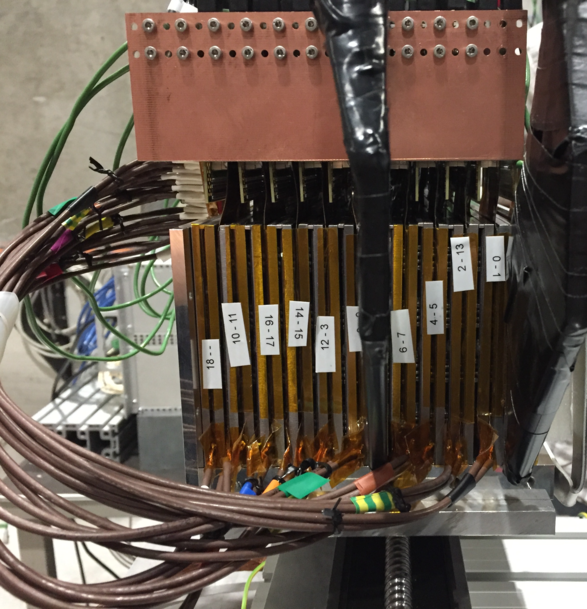}
\end{center}
\vspace{-5mm}
\caption{A side view of mini FoCal at SPS test beam including two plastic scintillation counters which were used to enhance electromagnetic shower signals. 
}
\label{fig:setup_SPS_side}
\end{figure}

\section{Data taking setup}
\label{sec:setup}

Data taking for the prototype detector was carried out at the SPS test beam facility and at the LHC (collision point ALICE) at CERN in 2018. 
The same readout system was used in all cases, as described above.
The data acquisition (DAQ) and control software was DATE~\cite{ALICE:2014tco} developed by the ALICE collaboration. 
The system was operated using a power distribution crate with a common trigger and clock distribution module~(CTGF\ v.6).
Reduced setups were used at the ELPH facility in Japan and in the PS test beam area to test single sensors to be used in the prototype.

For the beam trigger in the PS and SPS test beams, we used in total 4 different plastic scintillators in coincidence: 
(1)~of 10$\times$10~${\rm cm}^2$ area to detect the beam presence, (2)~of 4$\times$ 4~${\rm cm}^2$ area with 1 cm thickness, (3) and (4) of smaller size, which are placed horizontally and vertically to define together a 1$\times$1~${\rm cm}^2$ fiducial trigger in the center of the active detector area. 
In addition, we installed two plastic scintillation-counters~(EJ-200/212,  5~mm thickness, 93~mm x 279.5~mm) with Hamamatsu R5505-70 PMT~(1 inch, fine mesh) in between the layers (positions were depending on the data taking period) as shown in \Fig{fig:setup_SPS_side}. 
These trigger counters that were commissioned at the end of the SPS test beam were used to enhance electromagnetic shower signals in the ALICE P2 cavern.

At the PS, we used the T9 test beam line which provided secondary beams (with mixed electrons and hadrons) with a beam energy from 1 to 5~GeV. 
This beam time was mainly used to understand the calibration of the sensors and to commission the DAQ and trigger system. 

At the SPS, we used the H8 test beam line which provided 100--150 GeV positrons~\footnote{Except here, we refer to these high-energy positron as electron beam.}, 250 GeV electrons, and 180 GeV hadrons.
This beam time was mainly used to obtain the energy resolution of the prototype.
Additionally, by measuring the response of the trigger counters to 250~GeV electrons, and simulating the ratio of signals at 50 to 250~GeV, we have established the effective threshold for triggering on 50~GeV electrons, for later use in the measurements in ALICE.

For the data taking at LHC in ALICE, we installed the same setup used at the SPS test beam, including the EM shower trigger with the scintillation counters installed inside the prototype as described above.

\section{PS and SPS test beam results}
\label{sec:results_PS_PSP}

\subsection{MIP signal}
In preparation of the PS and SPS test beam experiments, we studied the performance of the sensors of the Mini FoCal at ELPH~(ELectron PHoton Science), Research Center, Tohoku University in Japan. 
We tested a single n-type 8$\times$8 silicon sensor of 320~$\mu {\rm m}$ thickness with the FPC shown in \Fig{fig:Sensor320FPC} using positron beams of about 900 MeV/$c$, where
the APV25 hybrid boards were directly mounted on the FPC. 
A clear MIP signal was observed for all pads which were tested in the position scan.


\begin{figure}[t!]
\begin{center}
 \includegraphics[width=60mm,clip]{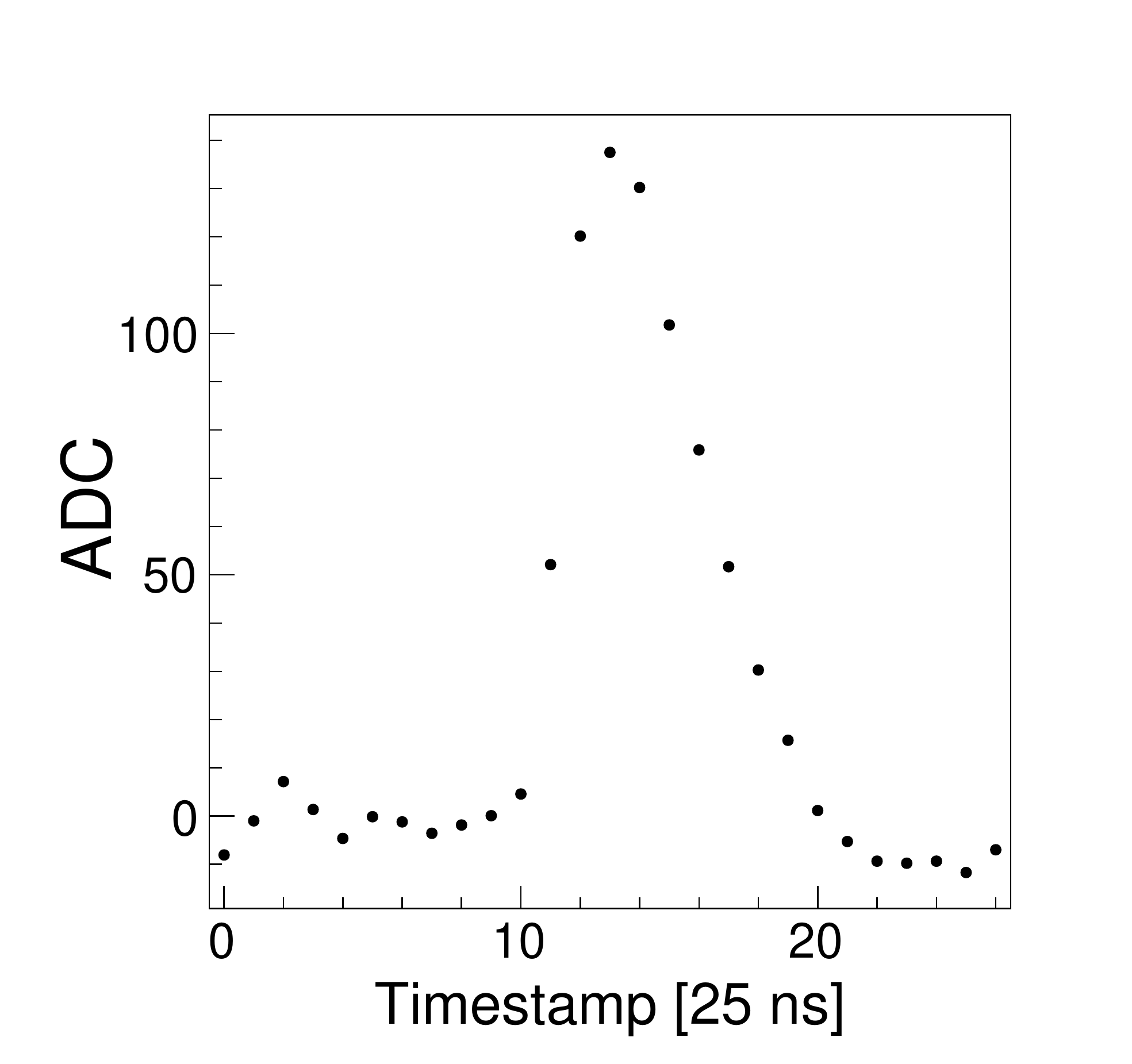}
\end{center}
\vspace{-5mm}
\caption{Example distribution of ADC samples of the APV readout in 27 timestamp bins with 25~ns increments as measured with a single sensor at the PS.}
\label{fig:signal}
\end{figure}

At the PS, a single sensor was exposed to 9~GeV hadrons without a tungsten layer in front in order to minimize secondary interactions. 
The sensor was moved to different positions to scan most of the pads and collect enough events to extract for each the MIP signal from the ADC distributions. This was used to determine the typical gain and the spread of the gains from channel to channel.
The APV readout samples the signal in every 25~ns interval. 
In the analysis, we reconstruct the full signal shape using a number of these samples. 
During the PS test, we made use of the 27 samples readout of the APV25. 
The delay of the signal was adjusted such that it arrives only after the seventh sampling interval, so that the first seven samples contain only the pedestal, as shown in \Fig{fig:signal}. 
In this way, the pedestals of every pad sensor were monitored individually and can be recomputed when needed to account for temperature, humidity or other variations. 

After pedestal subtraction, the common-mode noise~(CMN) was estimated from the individual channels and additionally subtracted. 
To do so, the response of all channels from lowest to highest values was sorted to ensure dead channels and signal channels were maximally separated.
The CMN was then estimated as the average of the three entries around the median of the sorted values.
The width of the pedestal was found to be about 11 ADC per channel dropping to 5.8--6.5 ADC after CMN subtraction.
To define the signal, we used a simple peak-finding algorithm, that requires a sequence of seven samples consisting of three subsequently increasing samples, followed by five decreasing values (including the maximum). In addition, the maximum ADC value is required to be the $8^{\mathrm{th}}$ sample or a later sample.
In case the criteria cannot be satisfied, the signal is discarded.
The difference between the ADC value of the largest signal in the sequence of selected samples and the pedestal value is used as the response of the detector.

\begin{figure}[t!]
\begin{center}
 \includegraphics[width=60mm,clip]{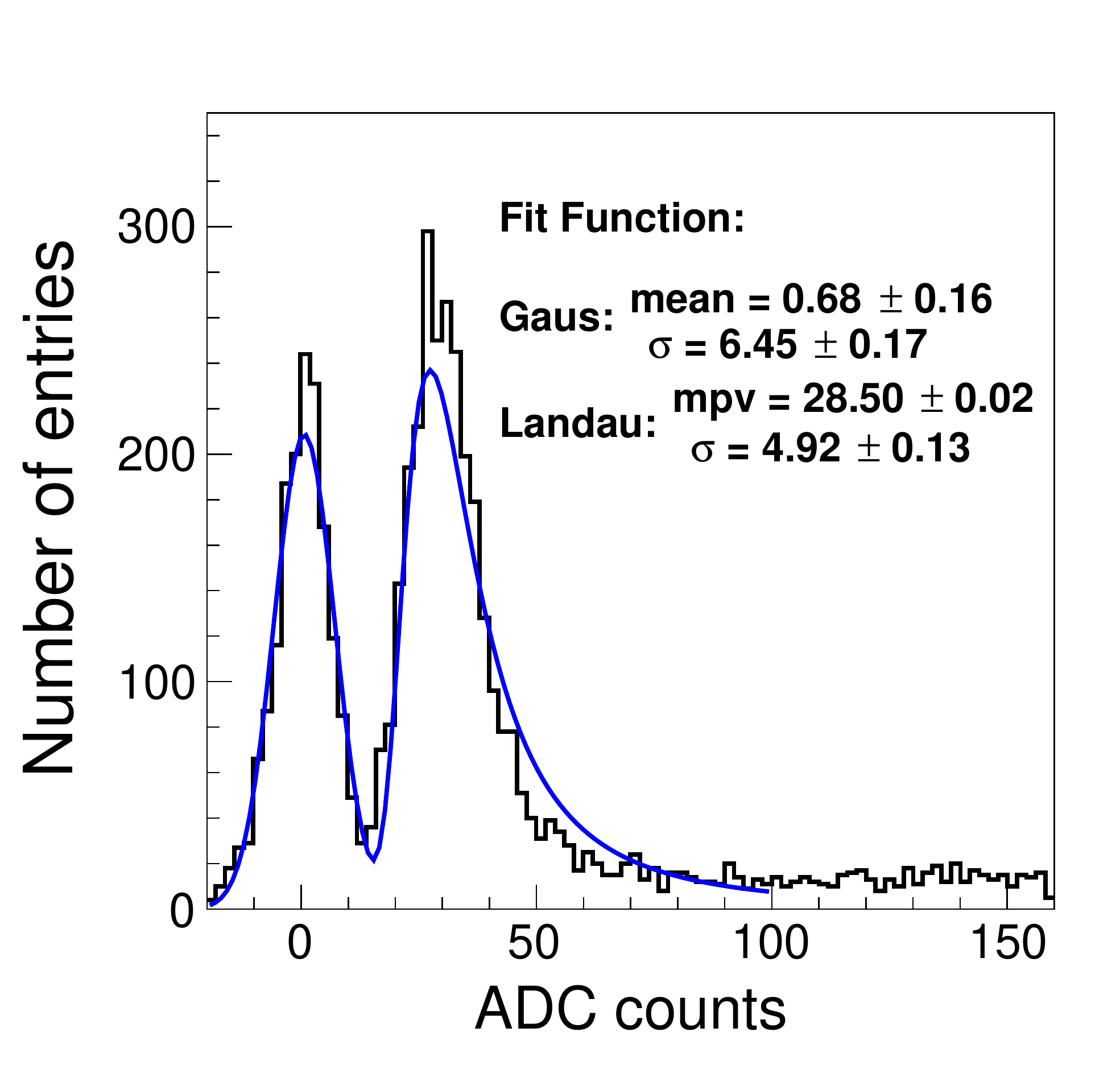}
 \includegraphics[width=60mm,clip]{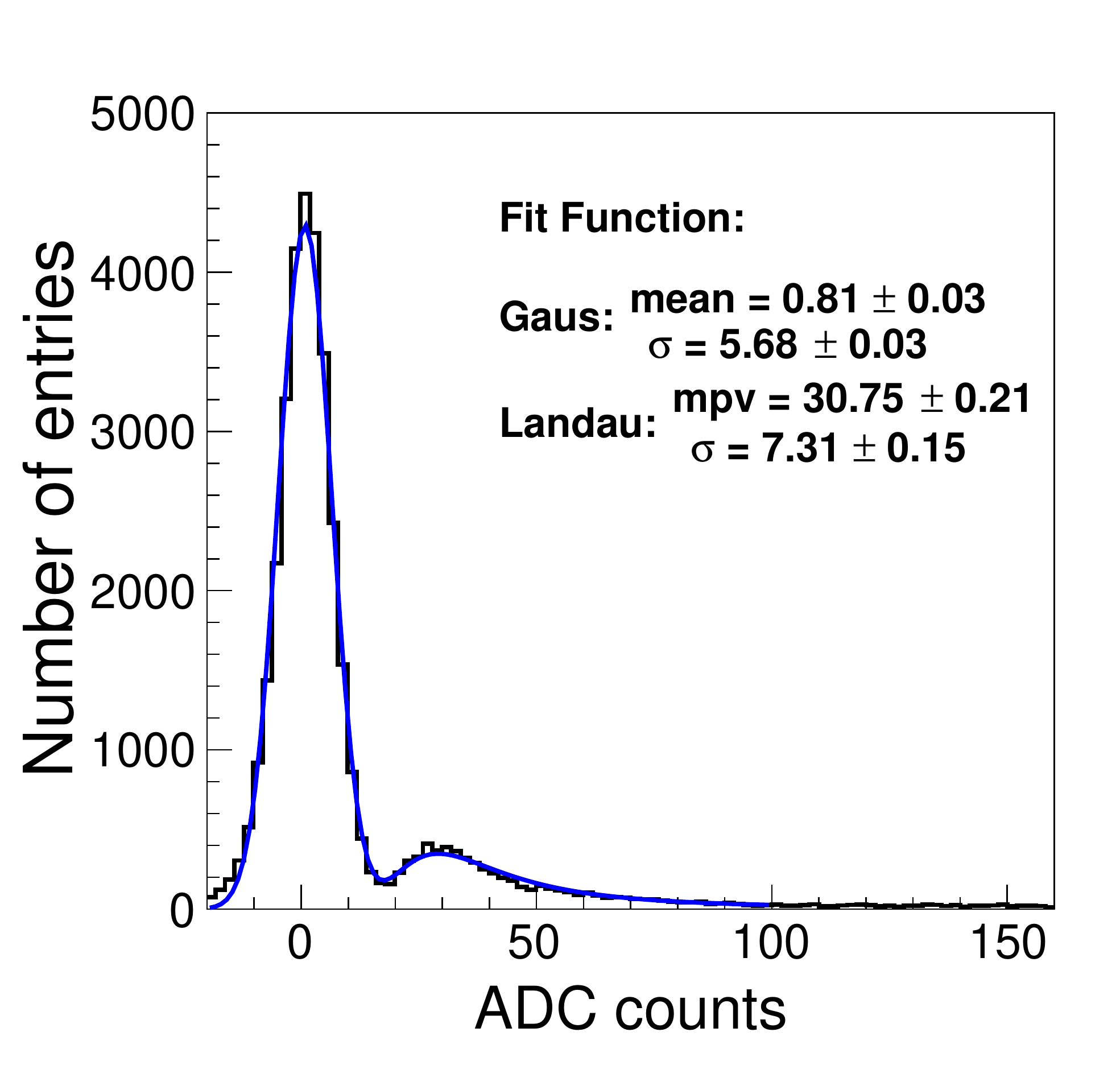}
\end{center}
\vspace{-5mm}
\caption{Distribution of ADC signals from single pads of one sensor after pedestal and common-mode noise subtraction in the triggered pad~(left) or in a $3 \times 3$ region around it~(right). Both distributions are fitted with a Gaussian and Landau fit functions for pedestal and signal regions, respectively.}
\label{fig:MIP}
\end{figure}

\begin{figure}[t!]
\begin{center}
 \includegraphics[width=60mm,clip]{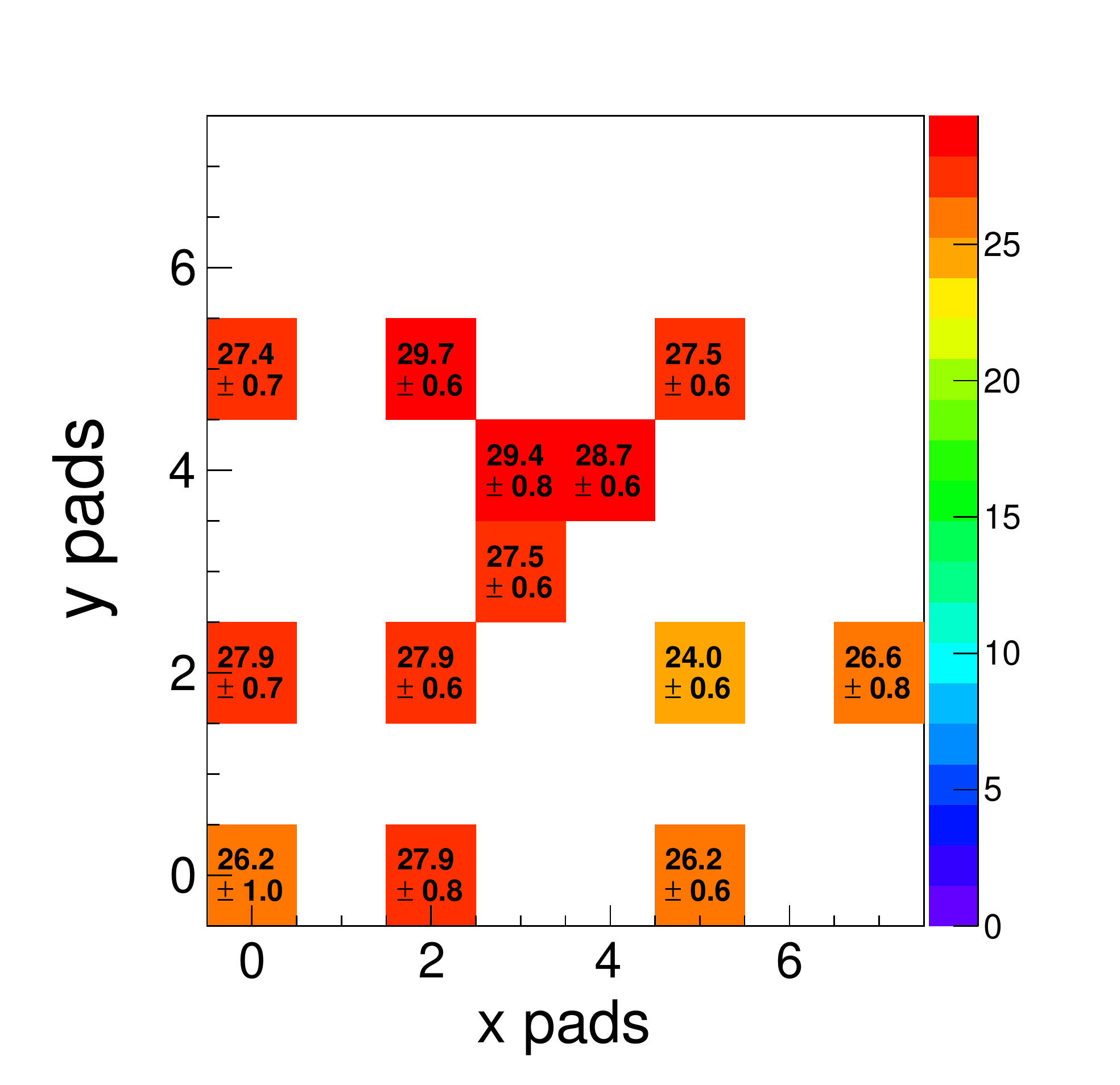}
\end{center}
\vspace{-5mm}
\caption{Reconstructed MIP peak position for different pads in a single sensor  without tungsten absorber in front. The most probably value of a Laundau distribution fitted to the response distributions is given, together with its statistical uncertainty.}
\label{fig:MIPmap}
\end{figure}

The scintillator setup covered approximately $1\times1~{\rm cm}^{2}$ area, corresponding to one pad of the silicon sensor. 
Therefore we focus on the MIP peak reconstruction in the center and the surrounding $3\times3$ pad area. 
The reconstructed MIP peak positions from single pads are shown in \Fig{fig:MIP} for the triggered~(central) pad or in the $3\times3$ region around it.
The ADC distribution of the MIP peak is fitted with a Landau distribution where the most probable value appears to be 28.5--30.75 with a width of 5--7.3 for the two selected regions. 

The MIP peak position from the simulation was extracted with the same method, and found to be about 92~keV per pad~(see \Fig{fig:muon}). 
By comparison with simulation it is found that one ADC count in the electronics corresponds to 2.9--3.2 keV in deposited energy. 
Therefore, we determine the noise level to be 16.8--20.8 keV depending on the channel in a single sensor.
In order to implement realistic noise in the detector response simulation, we extracted the signal-to-background ratio, calculated as the Landau MPV divided by the pedestal width, to be 4.4--5.3. 
In the following, the ``realistic'' simulation is one which includes the noise as well as dead channels, as opposed to the ``ideal'' simulation, which includes neither of the two.

In addition, we also extracted the MIP peak position for several geometrical locations on the 8x8 sensor, where the available data allowed us to fit the MIP peak position, as summarized in \Fig{fig:MIPmap}. 
We found about 10--15\% variation between the different locations, with no systematic dependence of the MIP peak position on the geometrical location.

\begin{figure}[t!]
\begin{center}
 \includegraphics[width=125mm,clip]{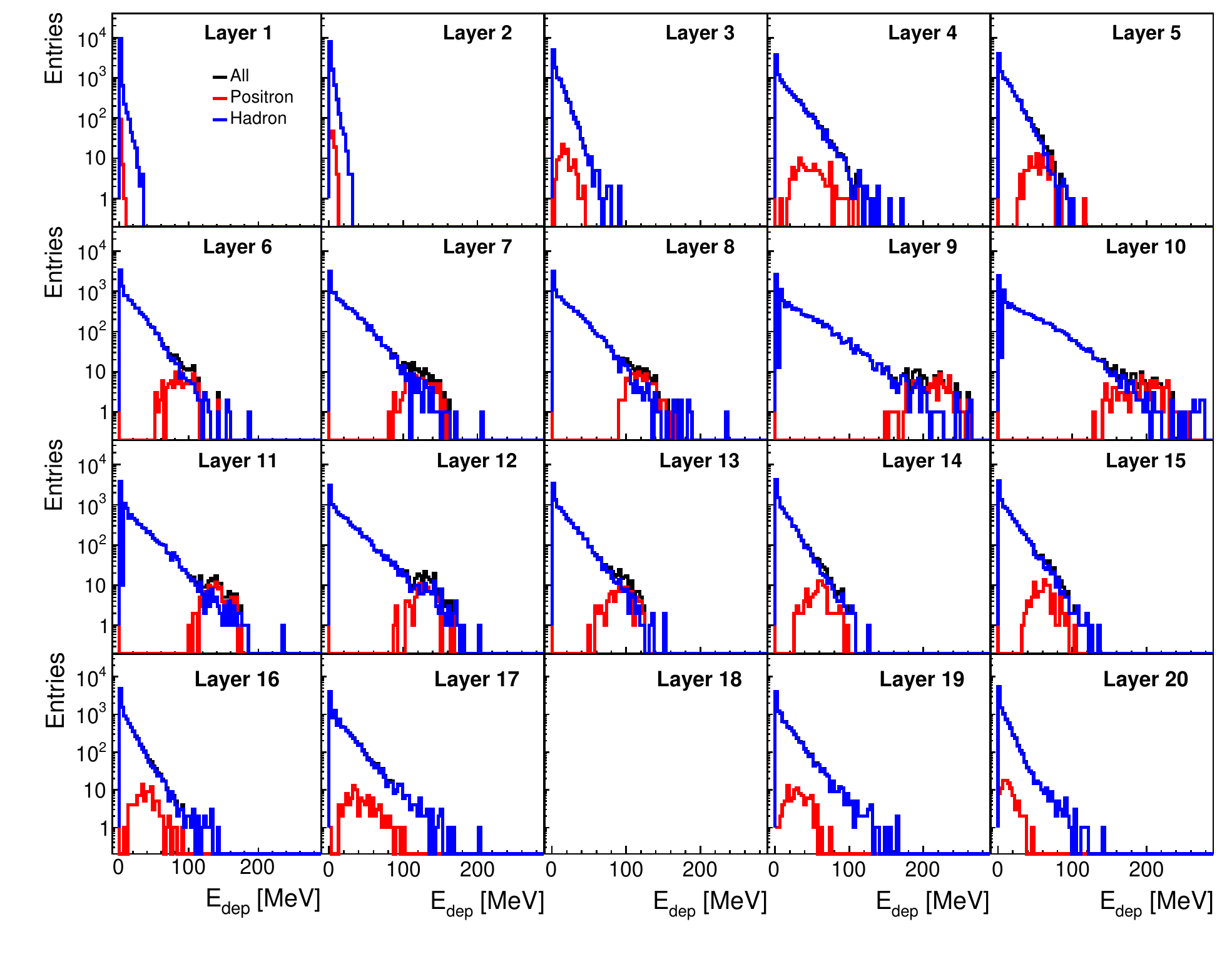}
\end{center}
\vspace{-5mm}
\caption{Distributions of deposited energy in the cluster in the Mini FoCal shown per layer at 150 GeV beam energy measured at the SPS. The red histograms represent the distributions applying the electron selection criteria, the blue histograms show the complement consisting mostly of hadrons.}
\label{fig:SelectionData}
\end{figure}

\begin{figure}[t!]
\begin{center}
 \includegraphics[width=125mm,clip]{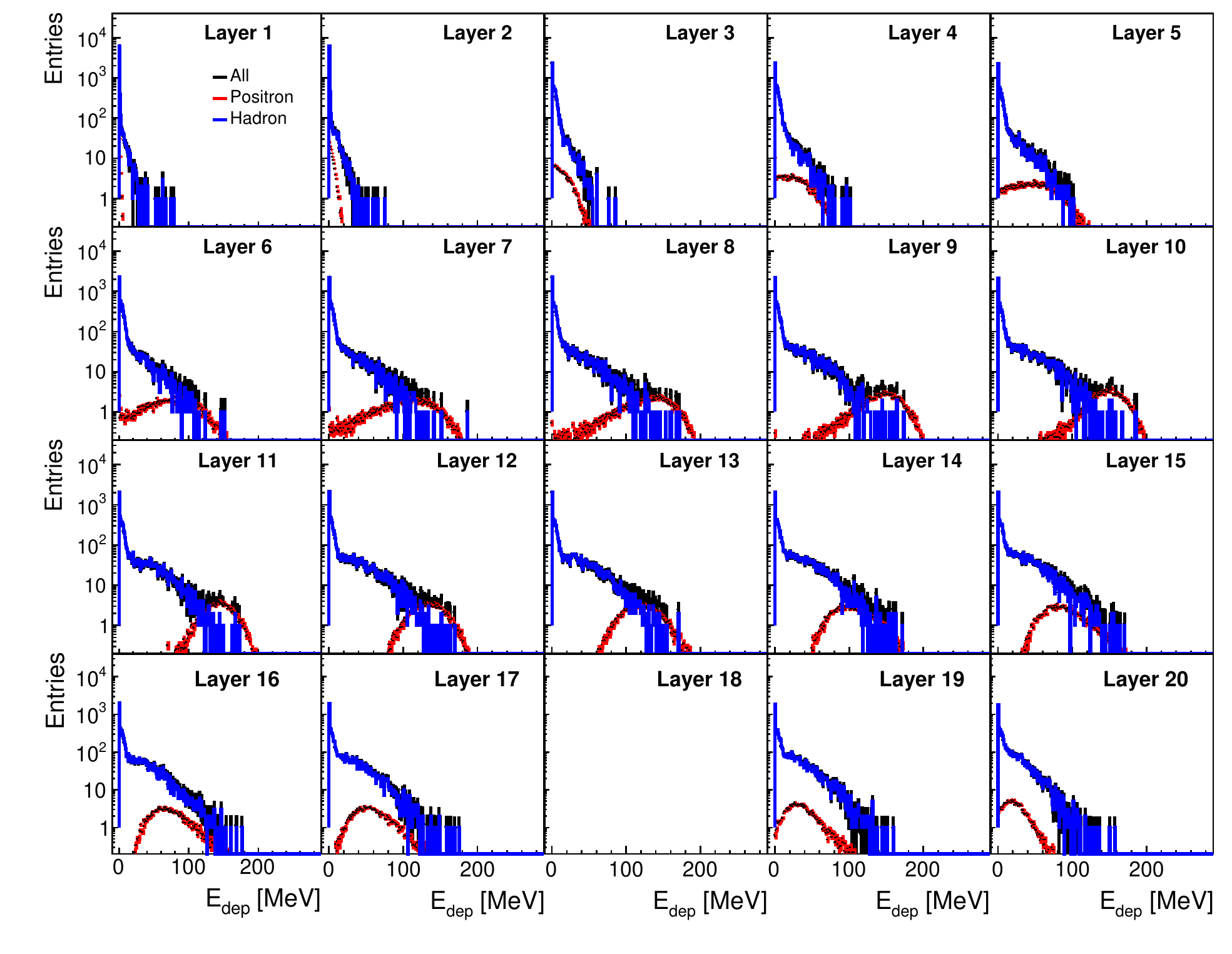}
\end{center}
\vspace{-5mm}
\caption{The simulated deposited energy in the cluster per layer at 150 GeV beam energy, obtained by mixing simulations with 2\% electrons and 98\% pions. The red histograms represent the distributions is the electron contributions, the blue histograms show the distributions of the pions.}
\label{fig:SelectionSim}
\end{figure}

\subsection{Energy resolution}
The energy resolution was obtained at the SPS, using beams of particles between 100 and 250~GeV.
As described above, we used the Mini FoCal with 20 layers of pad sensors interleaved with tungsten. 
Due to the limited range of the APV readout, we determined an attenuation of 1/18 for the first two layers and 1/180 for the remaining 18 layers prior to the testbeam. 
In this way, we were able to cover the dynamic range of the 100--250~GeV electron showers with the APV25 readout modules, while also maintaining good sensitivity to small signals in the first layers. 

The pedestal widths were similar to those observed in the PS testbeam, therefore we also adjusted the noise levels to $18\times$ or $180\times$ the previous value corresponding to the appropriate layers for the response in the simulation.

For the APV25 readout, we adjusted the readout time window so that the first three samples are always pedestals, followed by up to 20 samples of the signal from the amplifier-shaper output similarly as for the measurements at the PS. 
After pedestal and CMN subtraction, we combined the layers in order to identify the location of the maximum ADC value in the whole detector and hence the most likely impact position of the particle. 
In the cluster reconstruction we then defined a cluster as the 3x3 area around the maximum ADC value in the full detector. 

We collected sufficient data at energies of 150 and 250~GeV from a mixed beam of hadrons and electrons. 
In order to suppress the contribution from hadrons we applied a simple selection around the expected electromagnetic shower maxima, requiring high deposited energy~($E_{\rm dep} \geq 450~{\rm MeV}$ for the 150 GeV beam and $E_{\rm dep} \geq 500~{\rm MeV}$ for the 250 GeV beam) from the sum of layers 9, 10 and 11.
\Fig{fig:SelectionData} shows the layer-by-layer distribution for 150$~$GeV particles with and without the selection criteria, representing electron and hadron candidates, respectively.
The resulting distributions are qualitatively similar to simulations shown in \Fig{fig:SelectionSim}, where a sample of 10k pions and electrons with a ratio of 98\% to 2\% was used. 

\begin{figure}[t!]
\begin{center}
 \includegraphics[width=100mm,clip]{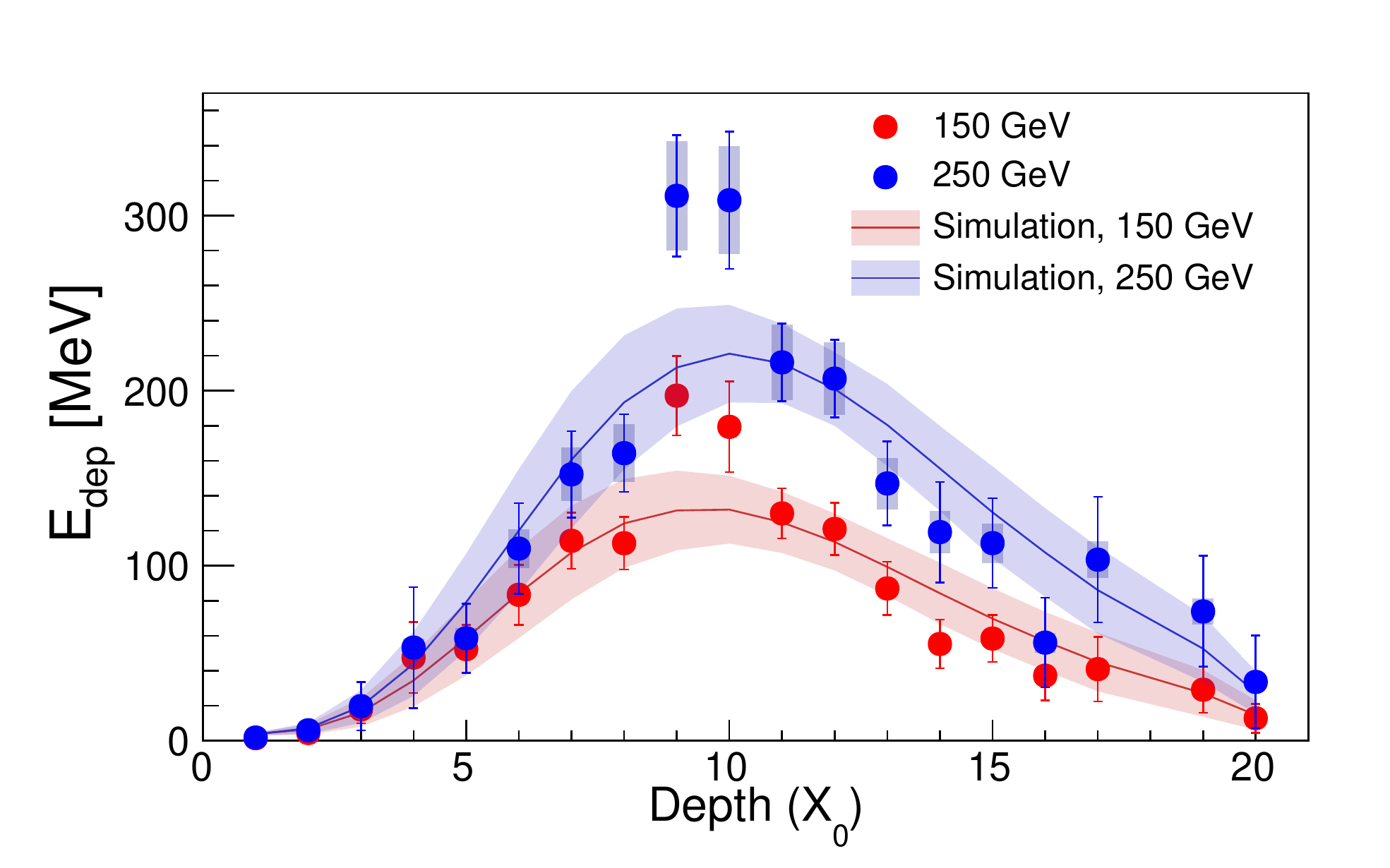}
\end{center}
\vspace{-5mm}
\caption{The longitudinal profile of the electromagnetic shower for selected candidates at 150 and 250 GeV. The points are showing the reconstructed data with their statistical and systematic uncertainties, while the lines are the mean responses from the simulation, the band corresponds to the sigma distribution with the realistic noise implementation.}
\label{fig:Longitudinal}
\end{figure}

In \Fig{fig:Longitudinal}, the longitudinal profile of the electromagnetic shower is compared to that obtained from the realistic simulation, for the selected electron candidates at 150 and 250 GeV.
The agreement between data and simulation is rather good, however some of the layers have higher or lower than expected responses, mostly seen in layers 9 and 10. 
This is due to (1) a variation in the APV25 gains, since we expect about $10\%$ difference between the readout electronics channels, and (2) the limited precision of the attenuation, which is expected to be about $10\%$ due to the component tolerance. 
The layers 9 and 10 shows the highest fluctuations and they were carefully studied not to originate from a selection bias. 
We found that the average ADC count without any selection is also about $30\%$ higher in these two layers.

\begin{figure}[t!]
\begin{center}
 \includegraphics[width=100mm,clip]{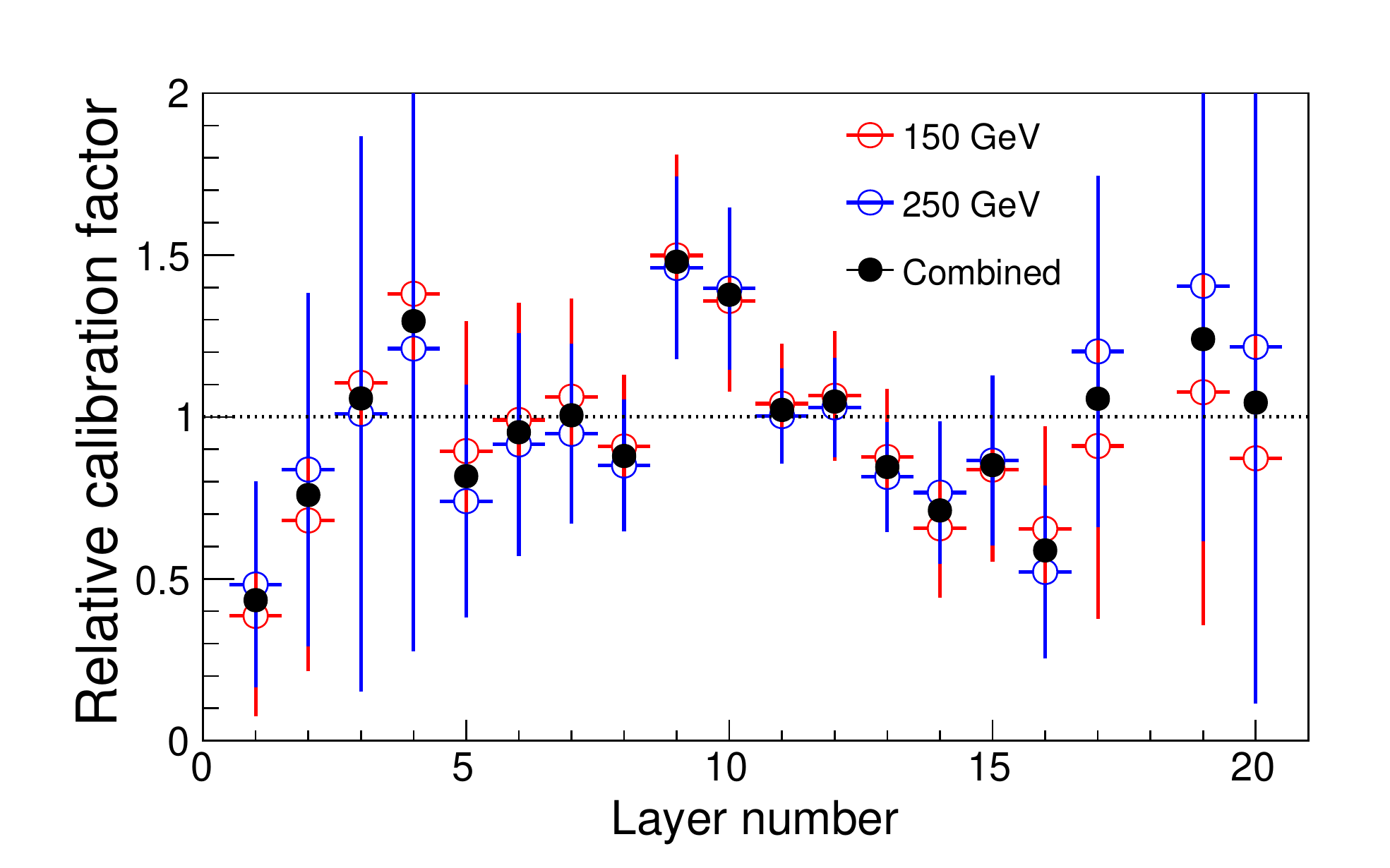}
\end{center}
\vspace{-5mm}
\caption{The calibration factors obtained for each layer by correcting the measured shower profile to the one from simulations.}
\label{fig:ReCalibration}
\end{figure}

\begin{figure}[t!]
\begin{center}
 \includegraphics[width=100mm,clip]{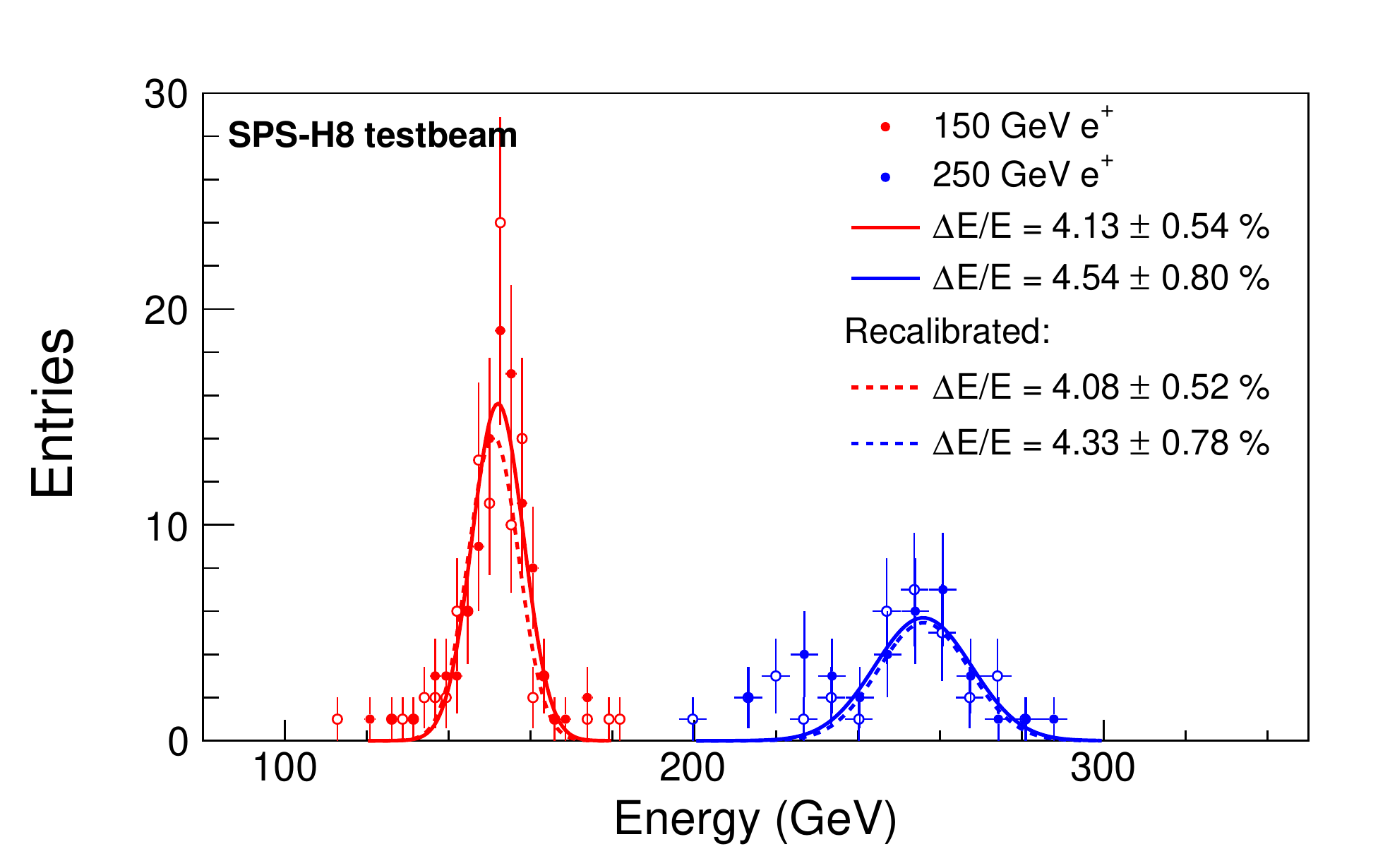}
\end{center}
\vspace{-5mm}
\caption{The total energy distribution for both 150 GeV and the 250 GeV beams with~(open markers) and without~(closed marker) relative calibration. Both distributions are fitted with Gaussian functions and the reconstructed resolution is around 4.3\% for both energies.}
\label{fig:Resolution}
\end{figure}

The absolute calibration of the detector response is determined from the expected beam energy and the average total signal at 150 GeV and then applied to both energies.
We then used the simulation for the relative calibration of the layers. 
The relative calibration factors are obtained as the ratio of signals in data to simulation for every layer by averaging the values between 150 and 250 GeV data, as shown by the solid black markers in \Fig{fig:ReCalibration}.
The total energy distributions with and without relative calibration for the 150 GeV and 250 data are shown in \Fig{fig:Resolution}. 
The energy resolution was obtained by fitting both distributions with a Gaussian function, resulting in a relative resolution of about 4.3\%.
The relative calibration leads only to a small improvement of the resolution. 

\begin{figure}[t!]
\begin{center}
 \includegraphics[width=100mm,clip]{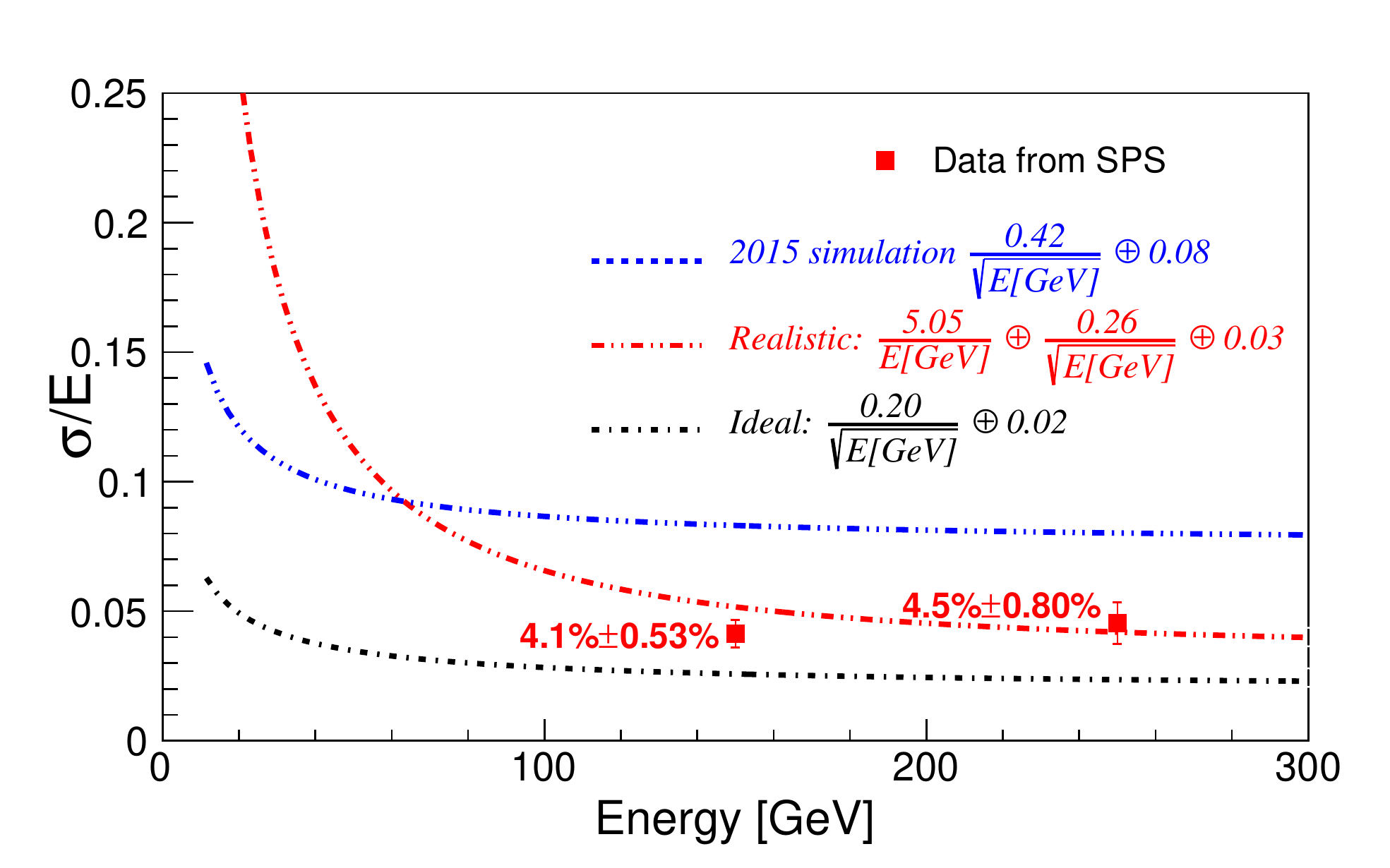}
\end{center}
\vspace{-5mm}
\caption{Energy resolution for the Mini FoCal detector as a function of beam energy compared to simulations. The blue curve represents the previous results from realistic simulations~\cite{Awes:2019vfi} reproducing data on lower energies, the black curve shows the ideal detector resolution, and the red curve represents simulations with realistic detector response for the detector studied here.}
\label{fig:ResolutionPlot}
\end{figure}

In \Fig{fig:ResolutionPlot}, the obtained resolution is compared to simulations, including those from the previous prototype~\cite{Awes:2019vfi} and the new prototype with the realistic noise.
For reference, we also show the result from ideal detector simulation without noise and without attenuation of the readout electronics, but using the same $3\times3$ clusterizer settings.

\begin{figure}[t!]
\begin{center}
 \includegraphics[width=60mm,clip]{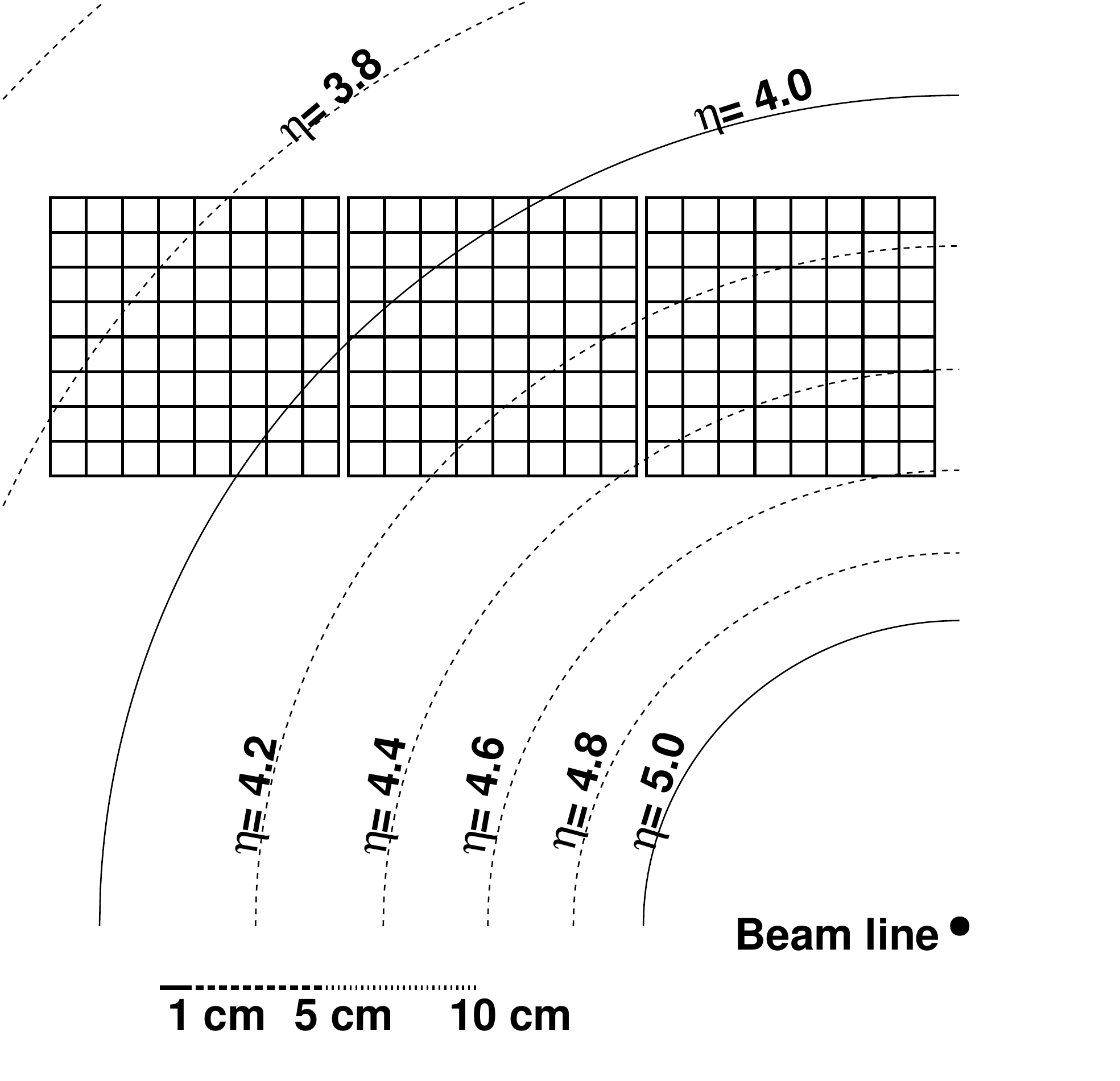}
\end{center}
\vspace{-5mm}
\caption{Acceptance of the prototype detector in pseudorapidity installed 7.5~m from the nominal interaction point.}
\label{fig:MiniFoCalGeometry}
\end{figure}

\section{Results obtained from proton--proton collisions at LHC}
\label{sec:alice}

The prototype detector was also installed in the ALICE cavern at point 2 at the LHC for a limited time from September to November in 2018. 
The detector was positioned close to the beam line, 7.5 meters away from the interaction point, covering pseudorapidities between $3.7<\eta<4.6$, as shown in \Fig{fig:MiniFoCalGeometry}.
As discussed above, we implemented a trigger on an electromagnetic shower of at least $\sim50$ GeV by installing two scintillators between the layers. 
Due to the close proximity of the detector to the LHC beam pipe, and the fact that the SRS system is not radiation hard, we experienced some losses of communication to the electronics.~\footnote{This failure has no consequences for the feasibility of operation of the final FoCal, because this part of the apparatus will be replaced by the HGCROC~\cite{Thienpont:2020wau} developed for CMS, which is known to have sufficient radiation tolerance.}
In the final measurements, we only could read out about half of the detector, namely the first 12 layers. 
Correspondingly, in the simulations we also did not use any information from the layers that were not read out.

\begin{figure}[t!]
\begin{center}
 \includegraphics[width=80mm,clip]{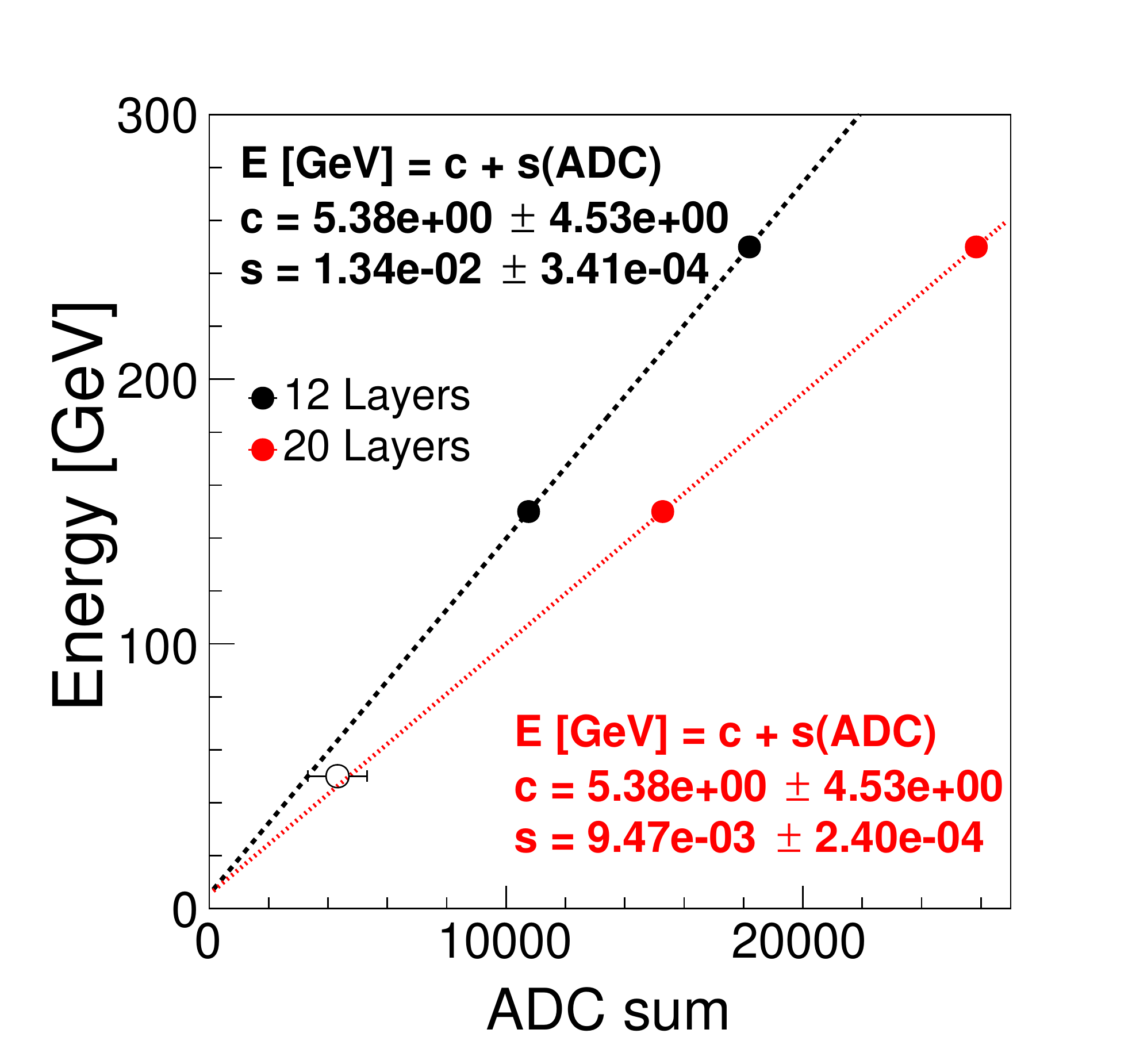}
\end{center}
\vspace{-5mm}
\caption{The energy per particle  as a function of the ADC sum per cluster in the Mini FoCal detector as it was in Point 2 (black) and in SPS (in red). The testbeam results were rescaled using the simulation from measured 20 layers to 12 working layers for energies 150 and 250 GeV. The 50 GeV point~(open marker) is obtained from the turn-on curve of the trigger. As this estimate relies on simulation it has significantly larger uncertainty.}
\label{fig:Calibration}
\end{figure}

\begin{figure}[t!]
\begin{center}
 \includegraphics[width=80mm,clip]{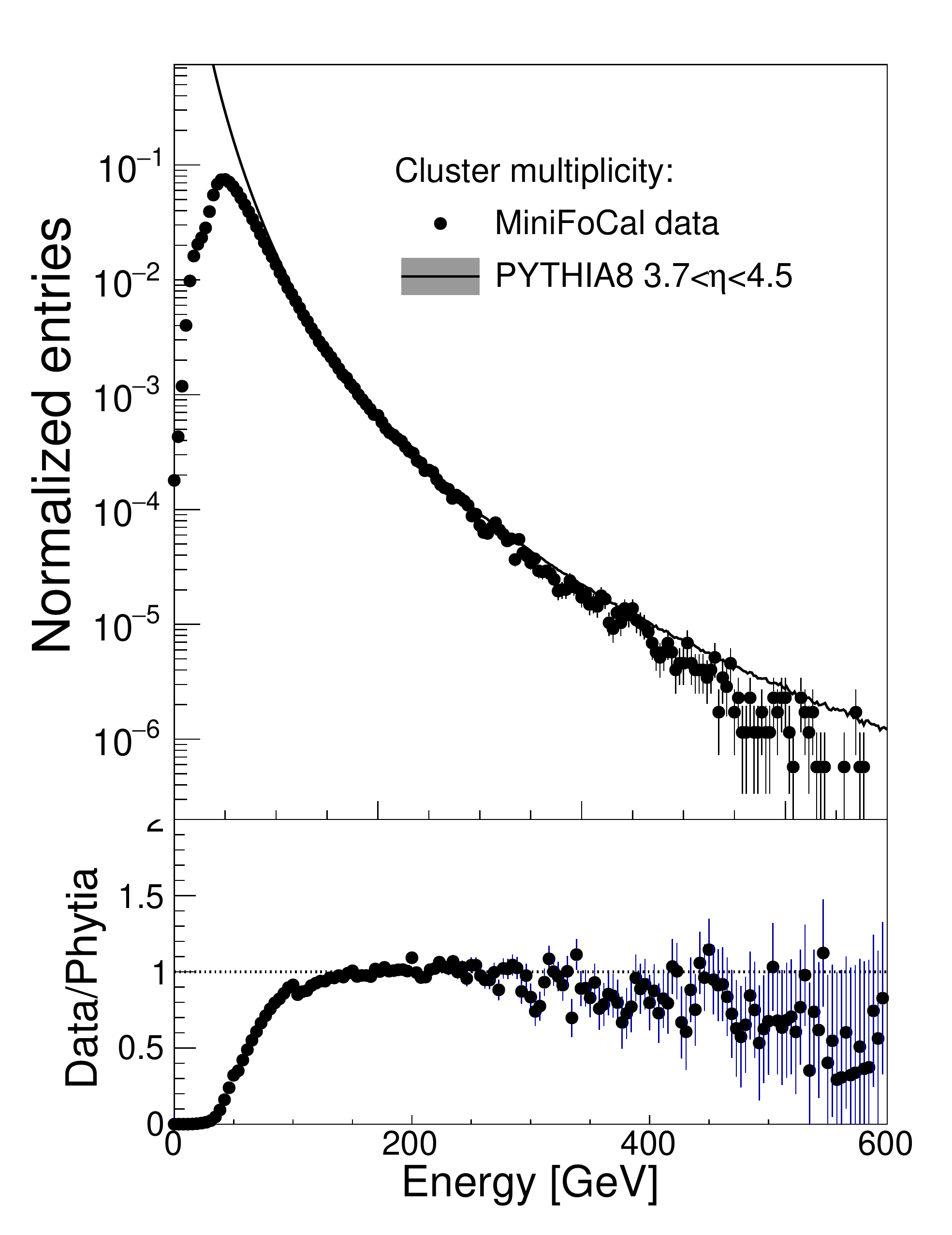}
\end{center}
\vspace{-5mm}
\caption{Distribution of reconstructed clusters in proton-proton collisions at $\sqrt{s}=$ 13 TeV as a function of cluster energy in the Mini FoCal in the $3.7<\eta<4.5$ pseudorapidity range compared to a PYTHIA8 simulation. The lower panel shows the ratio of data to simulation.}
\label{fig:ClusterMultiplicity}
\end{figure}

For the calibration, we used the measured 150 GeV and 250 GeV points from the SPS measurement. 
To account for the missing data of the layers for which the SRS system stopped functioning, simulation results were used to rescale the measured response. 
In addition, we tested the same scintillators in SPS to be used as a trigger at 50 GeV showers. 
Therefore, we use the expected turn-on curve of cluster energies around 50 GeV, but with a significantly larger uncertainty. The turn-on maximum is extracted from the energy spectra in \Fig{fig:ClusterMultiplicity} with the assumption that it is located at 50 GeV as seen in the figure.
Note that this first point is mostly for a sanity check to see, whether the electronics is responding according to the expectations, and it does not influence the overall calibration. 
The detector response in the testbeam was measured with 20 layers, and the response for only 12 working layers was rescaled using the simulation.
The detector response is consistent with a reasonably good linearity -- the readout electronics seems to behave linearly and there is apparently no strong effect from shower leakage on the linearity. 
We thus fit these points with a linear function in order to translate the ADC values of the cluster to the EM shower energy, as shown in \Fig{fig:Calibration}. 

\subsection{Cluster multiplicity}

\begin{figure}[t!]
\begin{center}
 \includegraphics[width=40mm,clip]{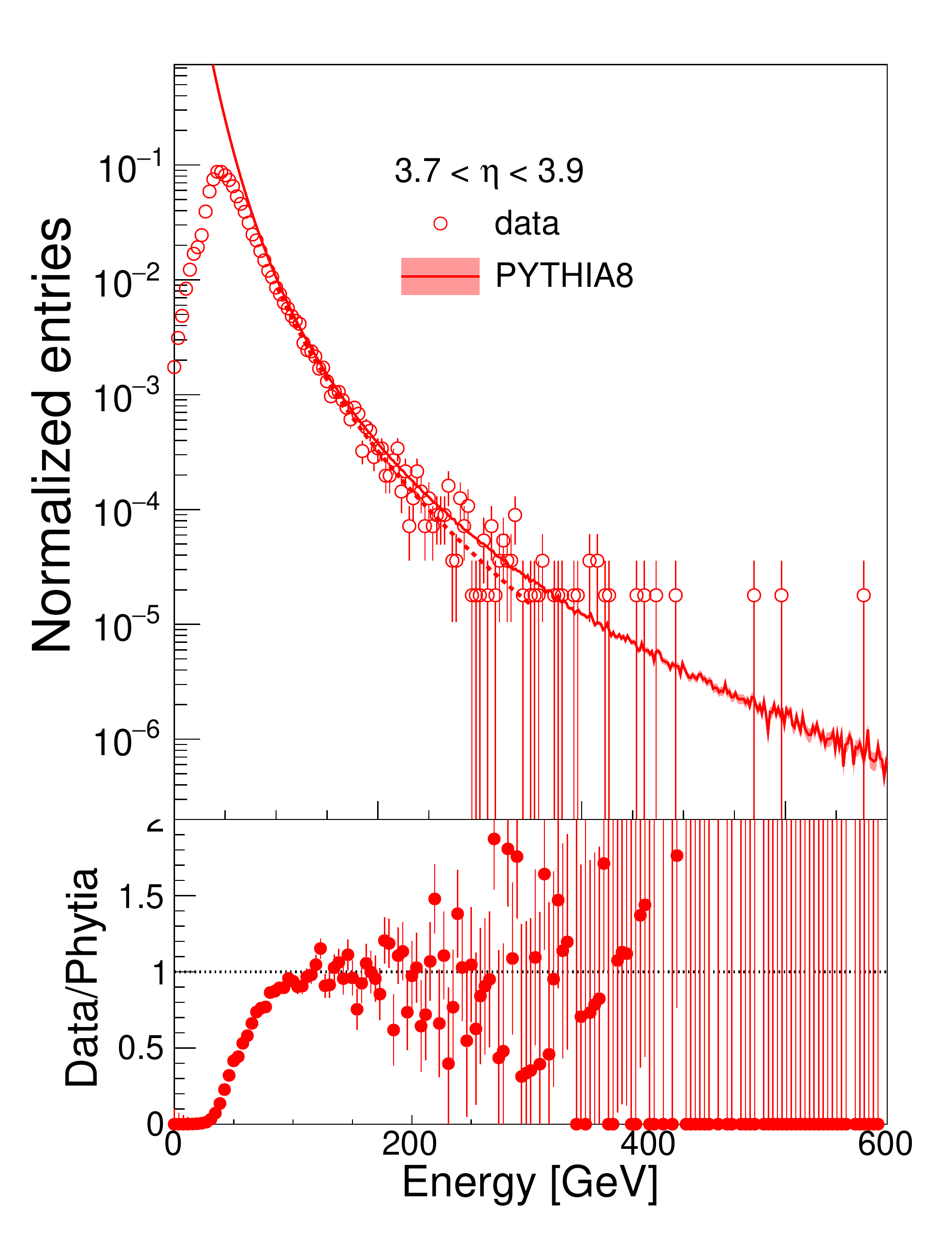}
 \includegraphics[width=40mm,clip]{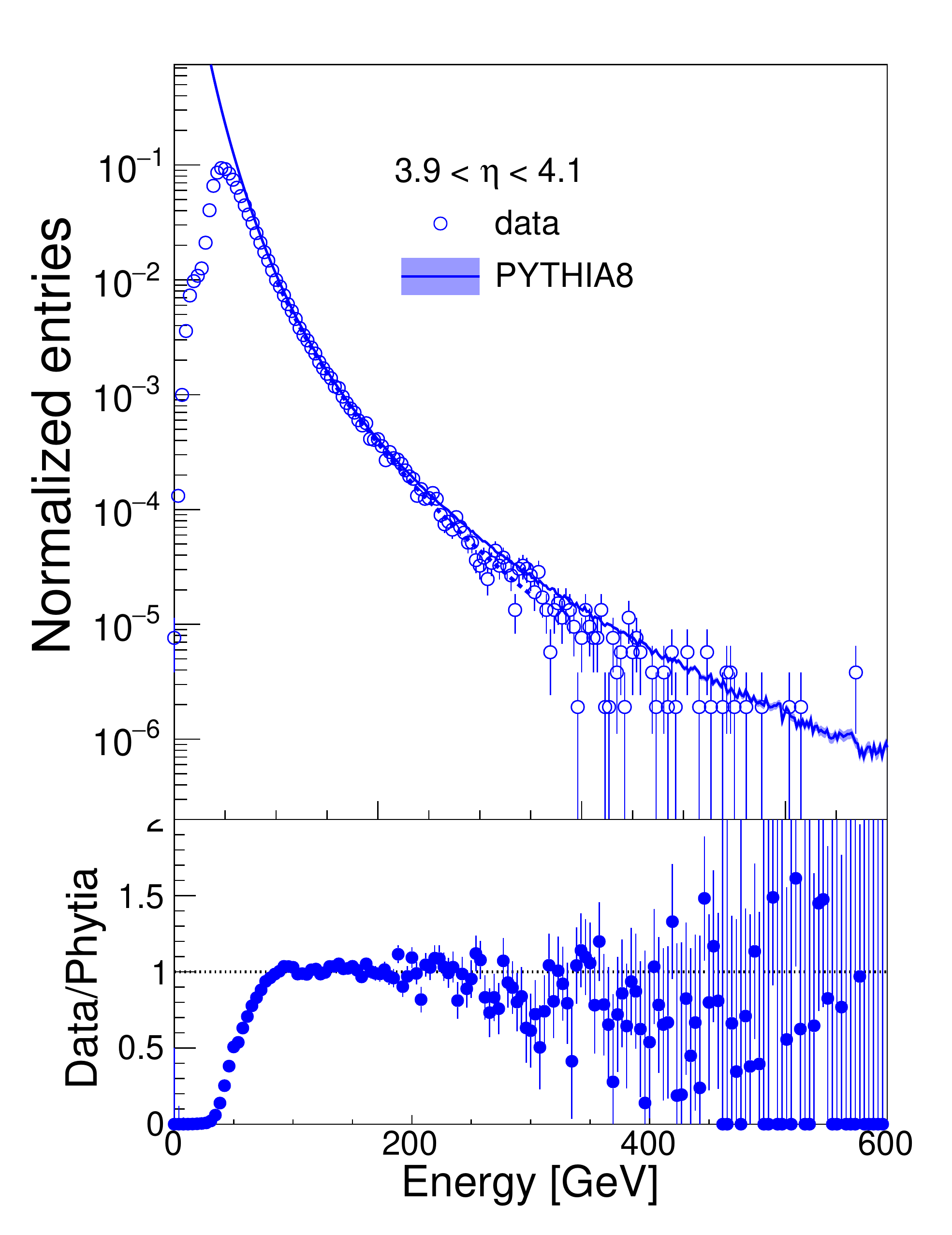}
 \includegraphics[width=40mm,clip]{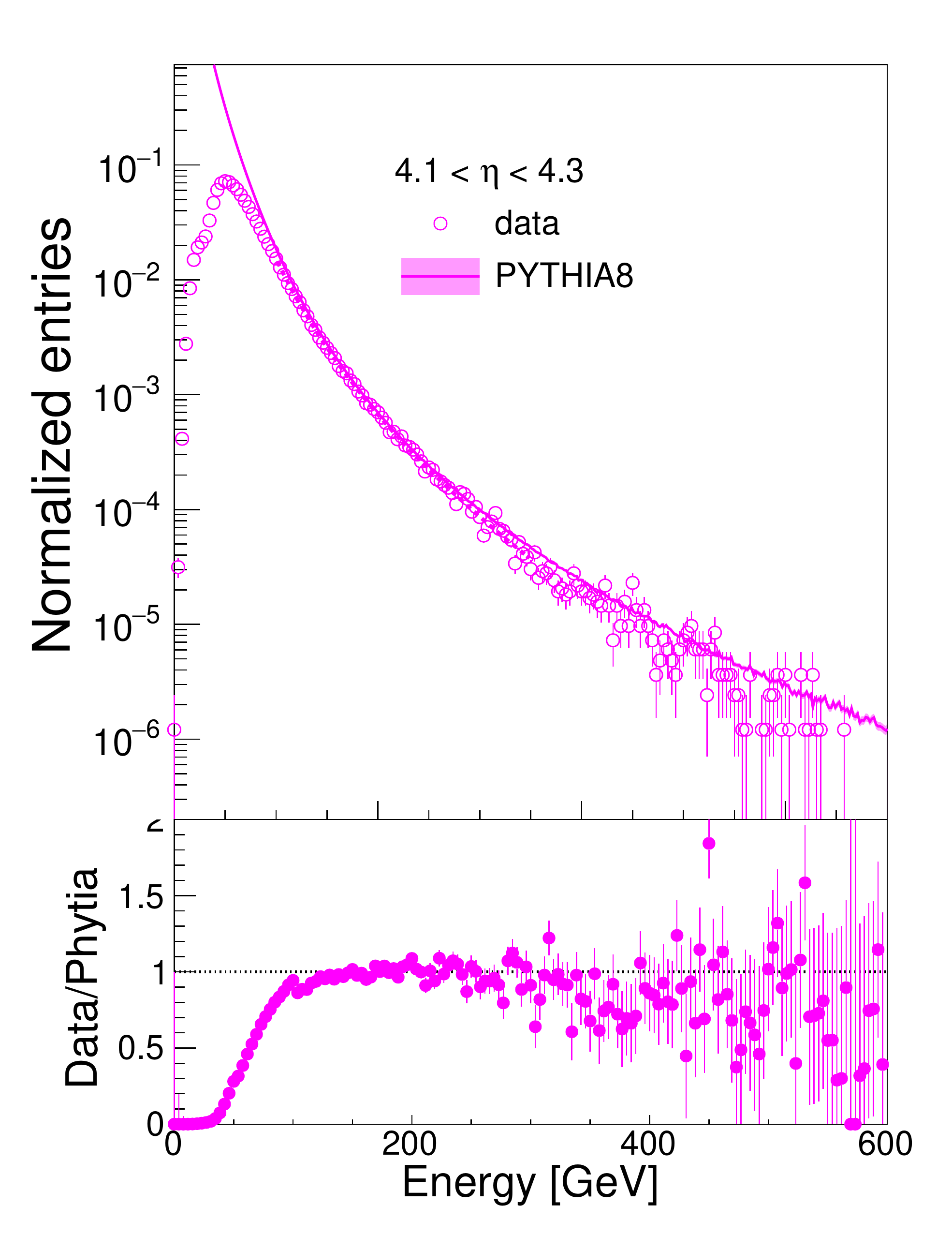}
 \includegraphics[width=40mm,clip]{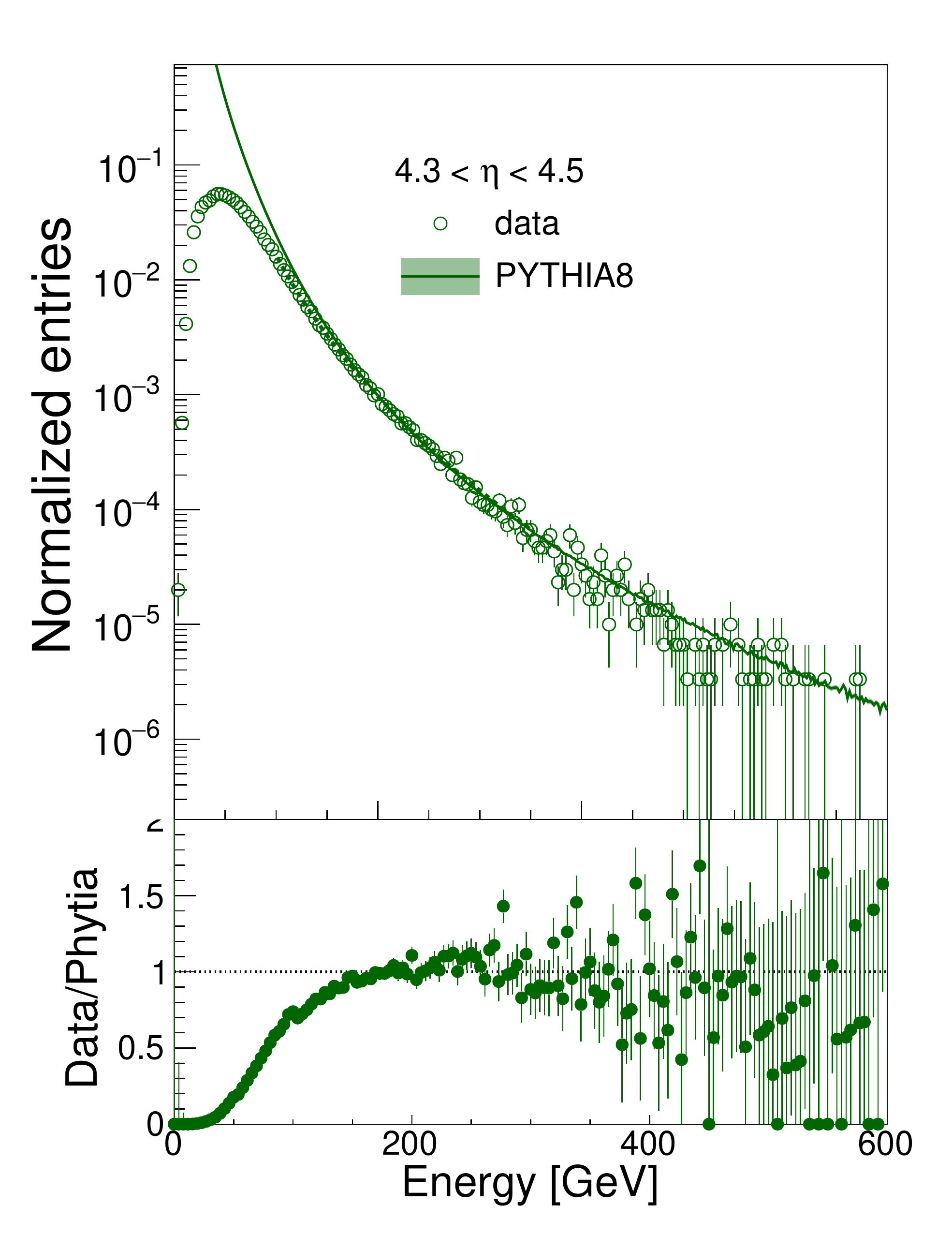}
\end{center}
\vspace{-5mm}
\caption{Distributions of reconstructed clusters in proton--proton collisions at $\sqrt{s}=$ 13 TeV as a function of cluster  energy in the Mini FoCal in four different pseudorapidity ranges compared to PYTHIA8 simulations. The lower panels show the ratios of data to simulation.}
\label{fig:ClusterMultiplicityRapidity}
\end{figure}

\Fig{fig:ClusterMultiplicity} shows the reconstructed cluster spectrum in proton--proton collisions at $\sqrt{s}=$ 13 TeV as a function of the cluster energy. 
The reconstruction was limited to within the $3.7<\eta<4.5$ pseudorapidity window. 
The results are compared to a PYTHIA8 particle-level simulation of the same collisions. 
Since we cannot identify the sources of the measured clusters in the prototype, we obtain the inclusive distribution of photons and electrons in the Mini FoCal acceptance from the PYTHIA8 simulation. The comparison between the simulation and data is also shown with a ratio in the lower panel of \Fig{fig:ClusterMultiplicity}. 
In this comparison we have normalized the PYTHIA8 distribution to match the data for high cluster energy. 
We have repeated the analysis of both data and simulation in four different rapidity windows, and the results are shown in \Fig{fig:ClusterMultiplicityRapidity}. 
Data and simulation appear to agree reasonable well in shape for energies significantly above the trigger threshold.

\begin{figure}[t!]
\begin{center}
 \includegraphics[width=60mm,clip]{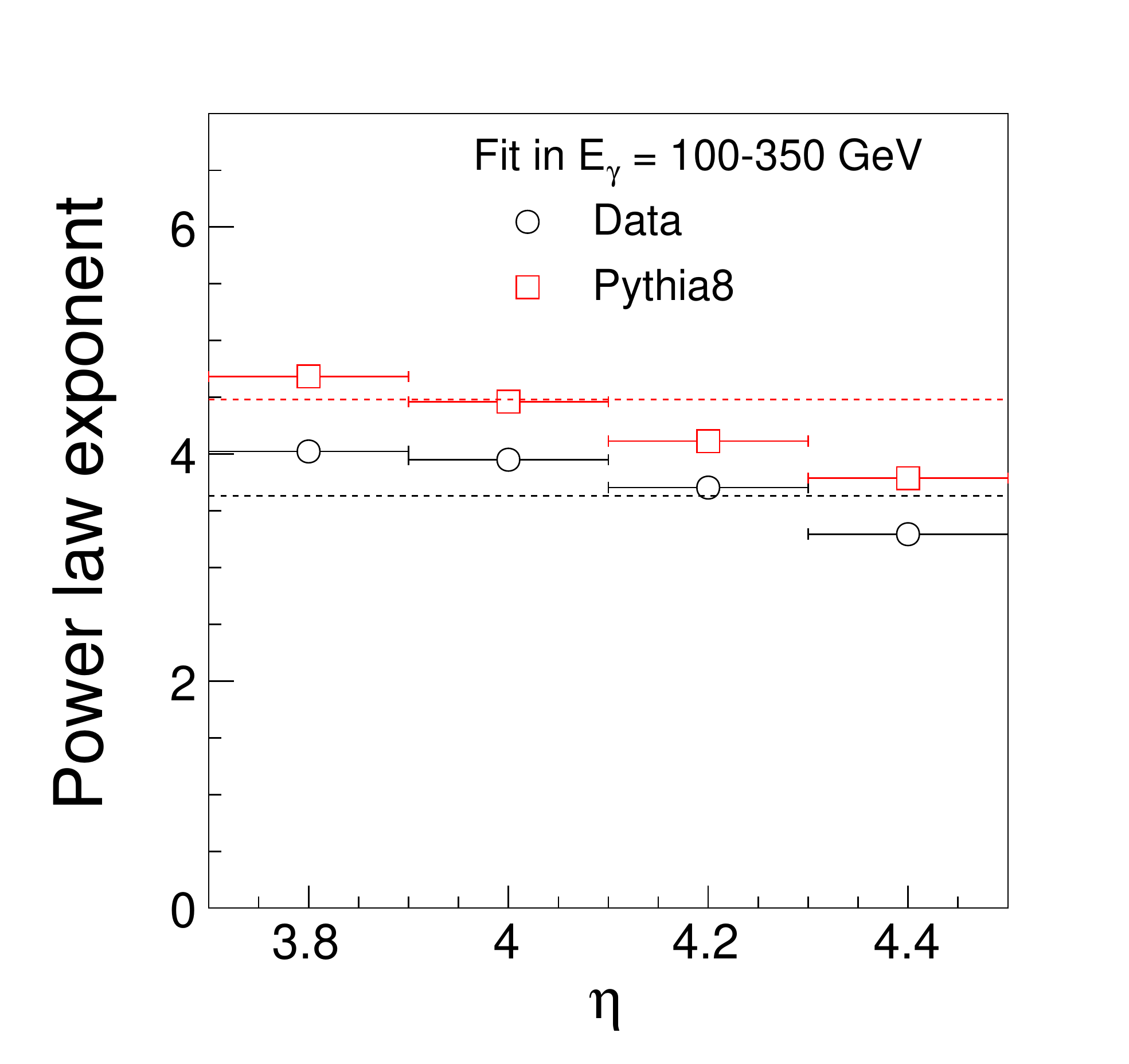}
\end{center}
\vspace{-5mm}
\caption{Power law exponent $n$ from fits to the cluster spectra from the data and the PYTHIA8 simulation as a function of pseudorapidity. The dotted lines show the average values over the full pseudorapidity region.}
\label{fig:FitSlope}
\end{figure}

We fit both the data and simulation results with a modified power-law function, $f(E) = A\cdot e^{b_1 + b_2E} E^{-n}$ function above 100 GeV where the trigger implemented in the Mini FoCal is already fully efficient. 
For consistency, the same region is also used in the PYTHIA8 simulated data. 
The results from data and simulation for the exponent $n$ of the power law are compared in \Fig{fig:FitSlope} and show similar behaviour as a function of pseudorapidity, however, we observe systematically larger values of $n$ in the simulated data, corresponding to slightly steeper spectra. This difference agrees qualitatively with expectations, because the simulations used here do not include any resolution effects that are present in the data and would broaden the distributions there.

\subsection{Invariant mass}

\begin{figure}[t!]
 \includegraphics[width=61mm,clip]{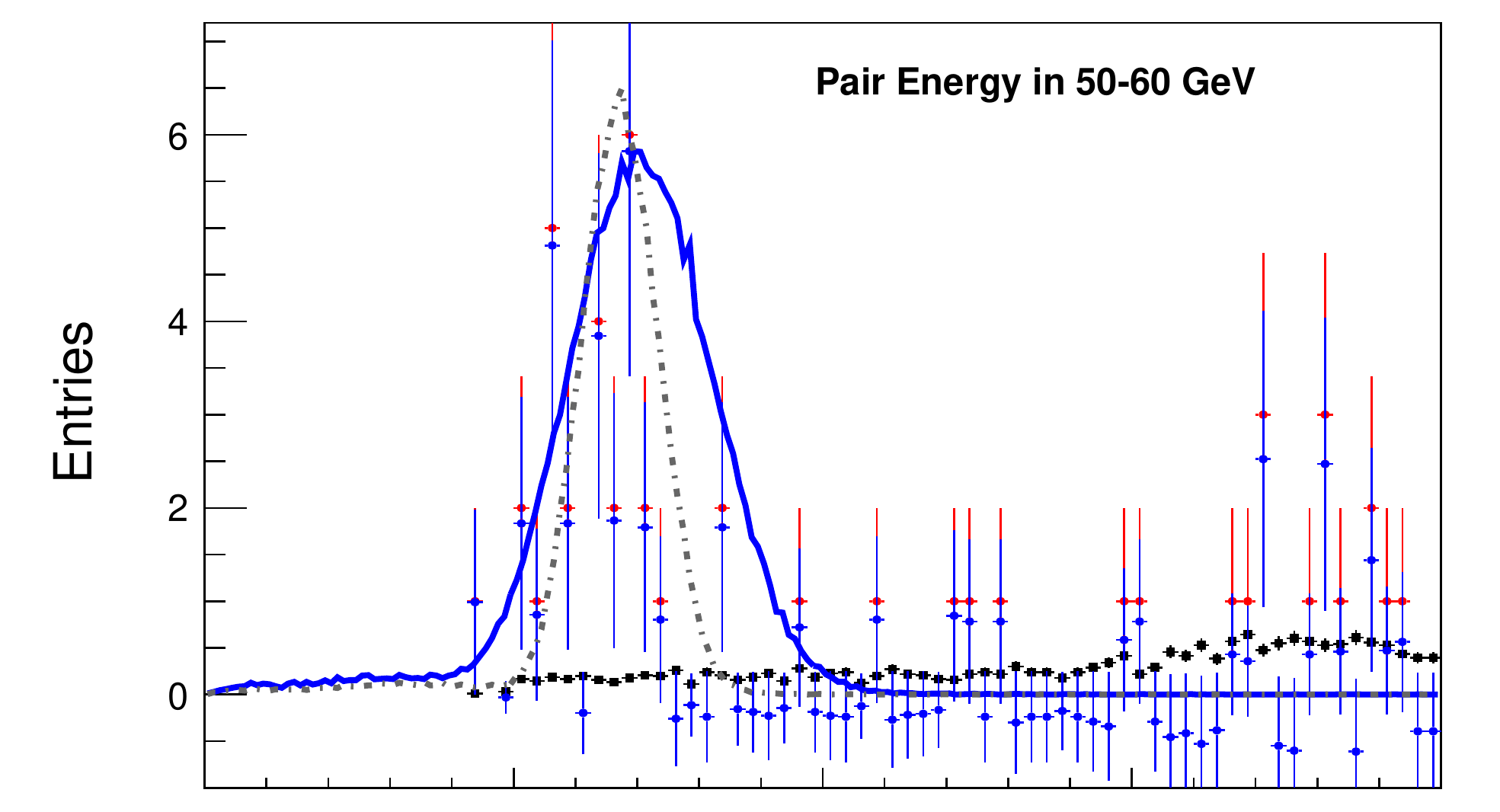}
 \includegraphics[width=58mm,clip]{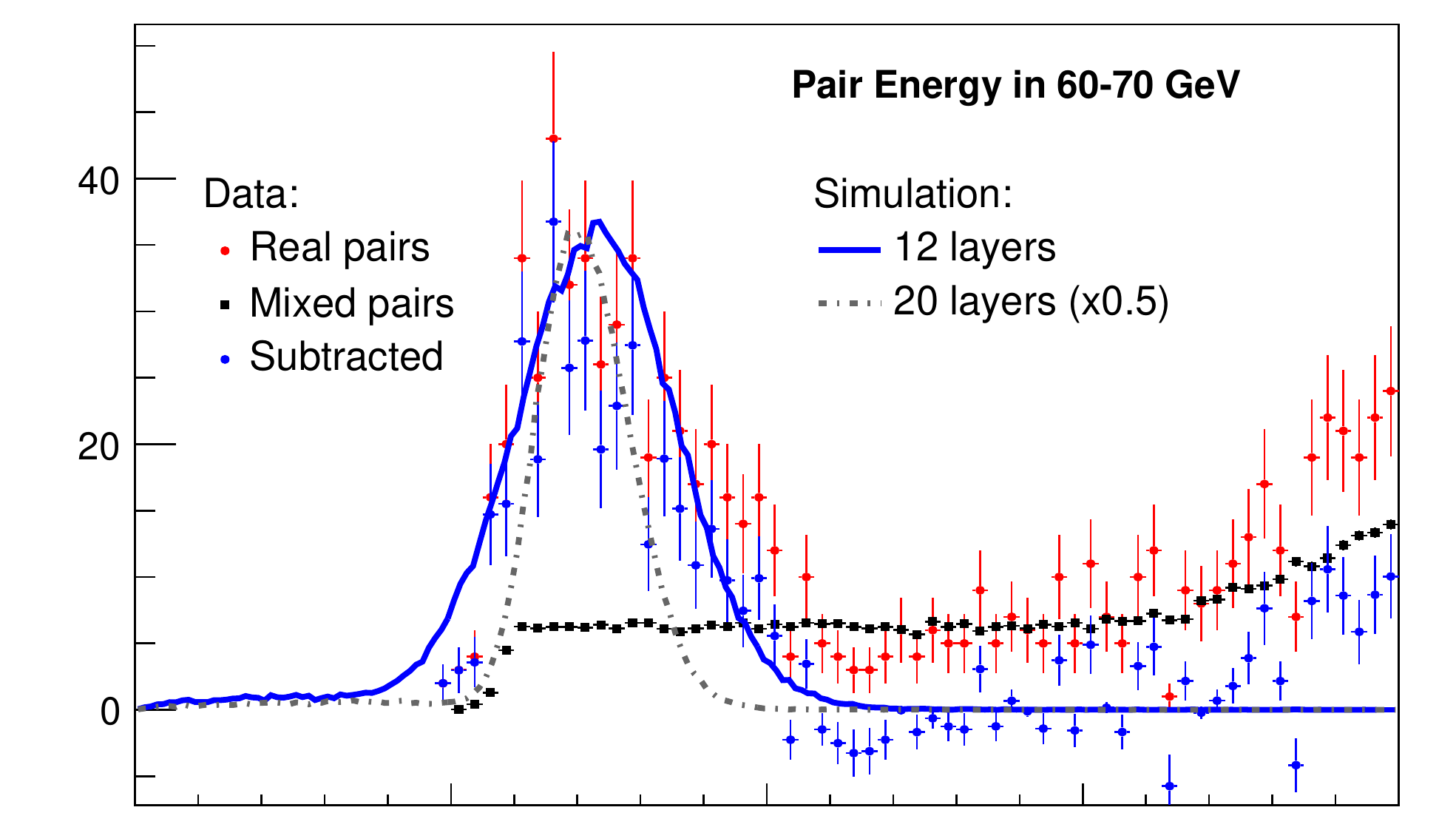}
 
 \includegraphics[width=61mm,clip]{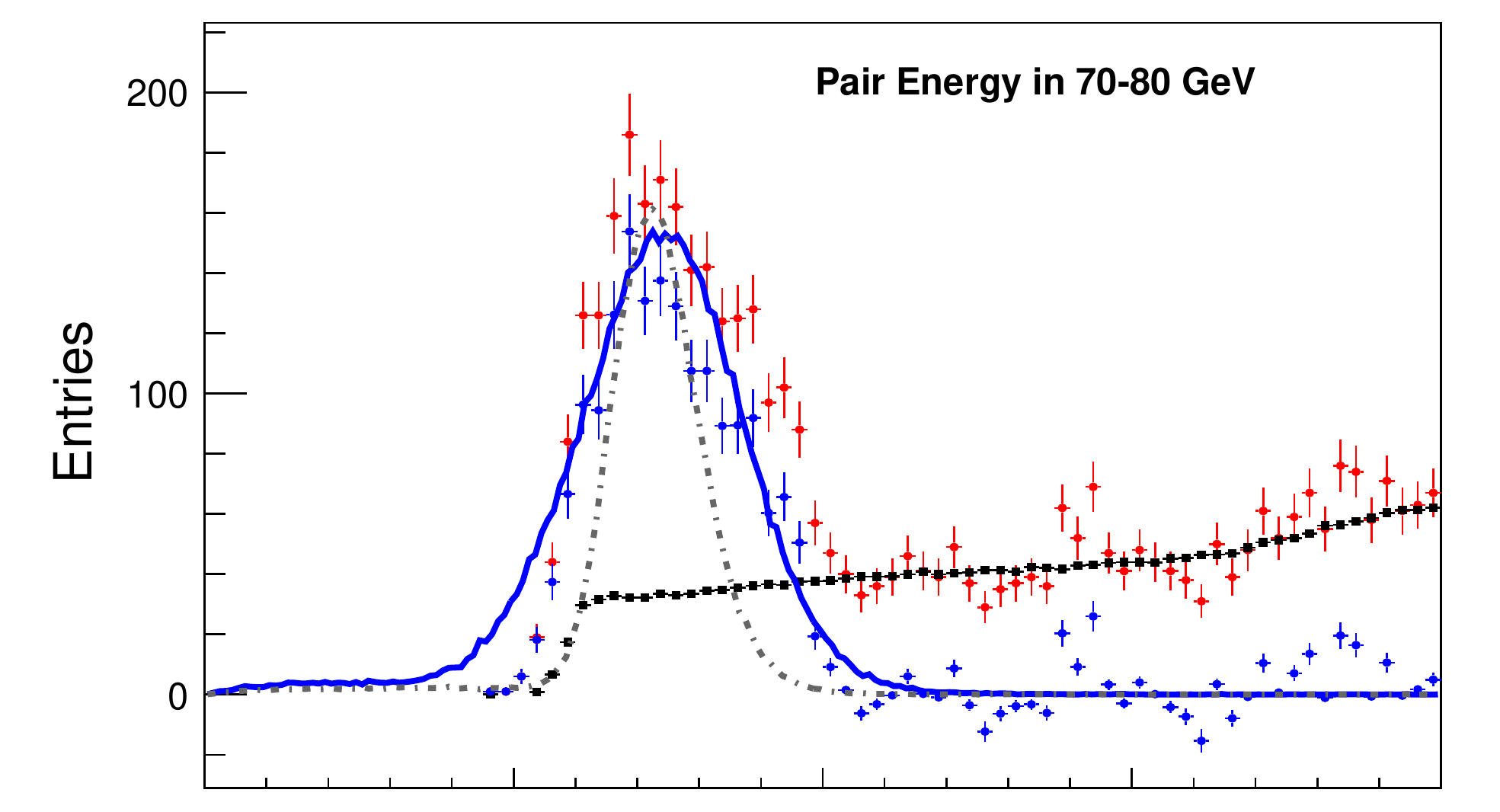}
 \includegraphics[width=58mm,clip]{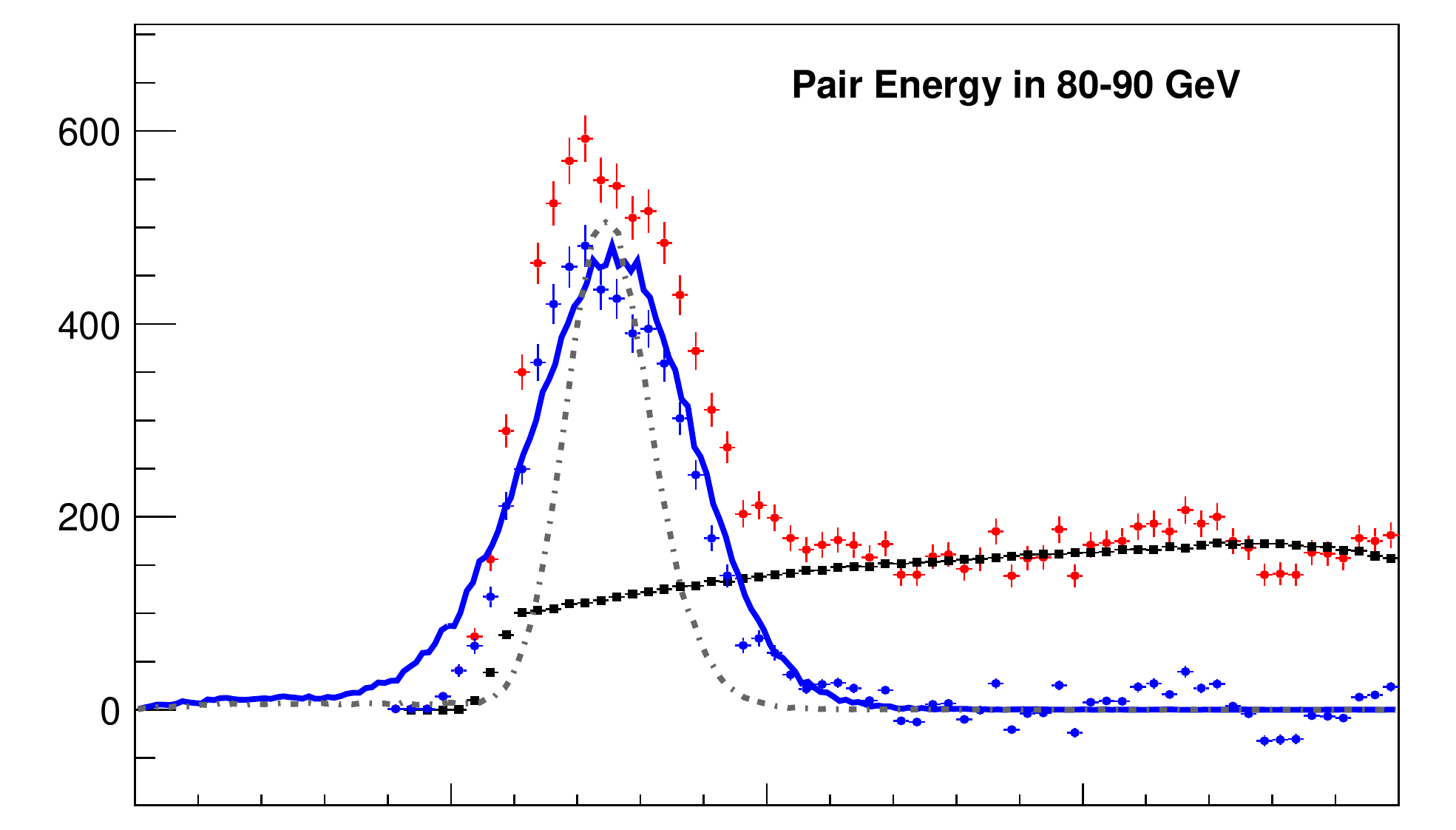}
 
 \includegraphics[width=61mm,clip]{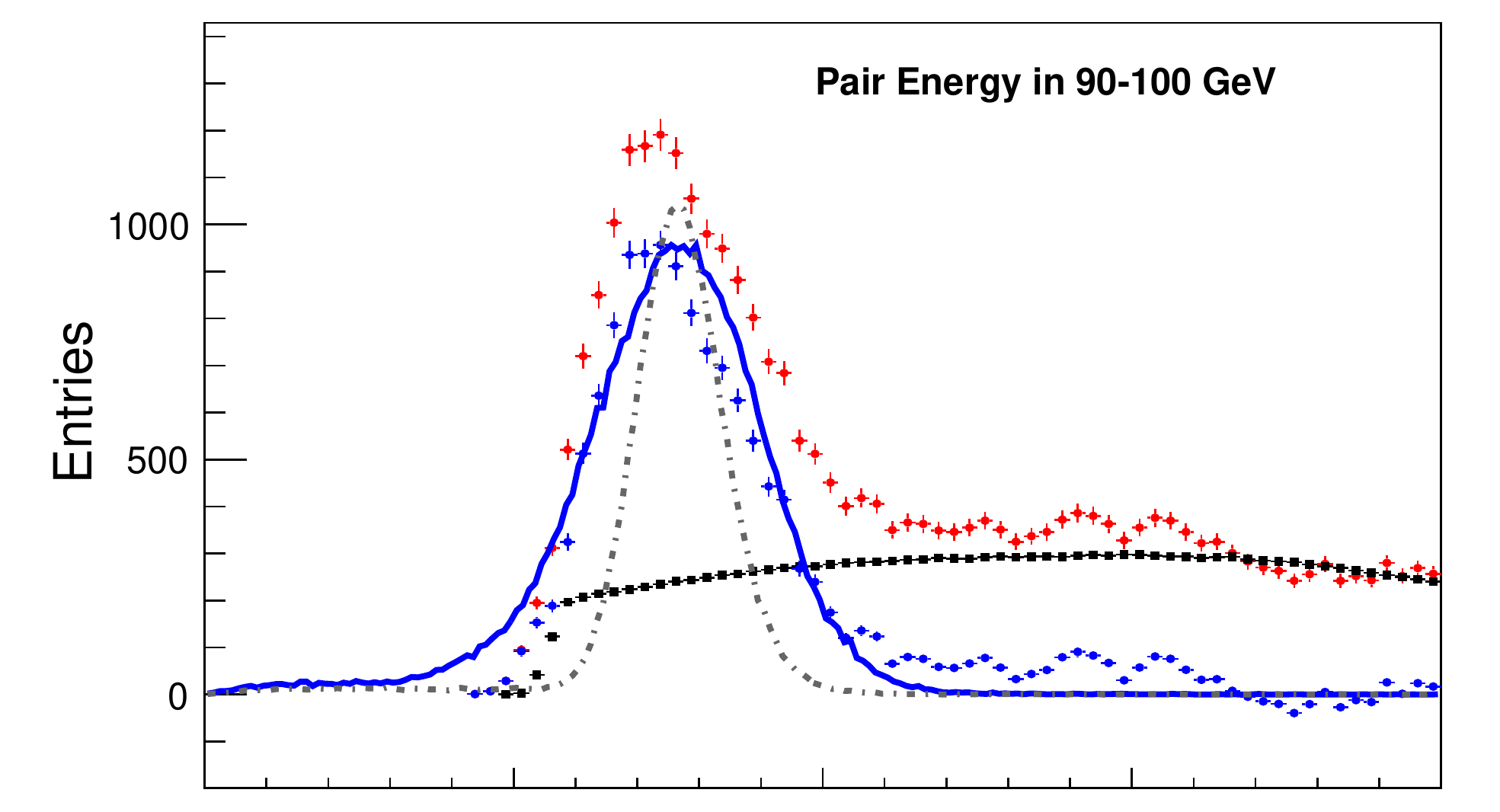}
 \includegraphics[width=58mm,clip]{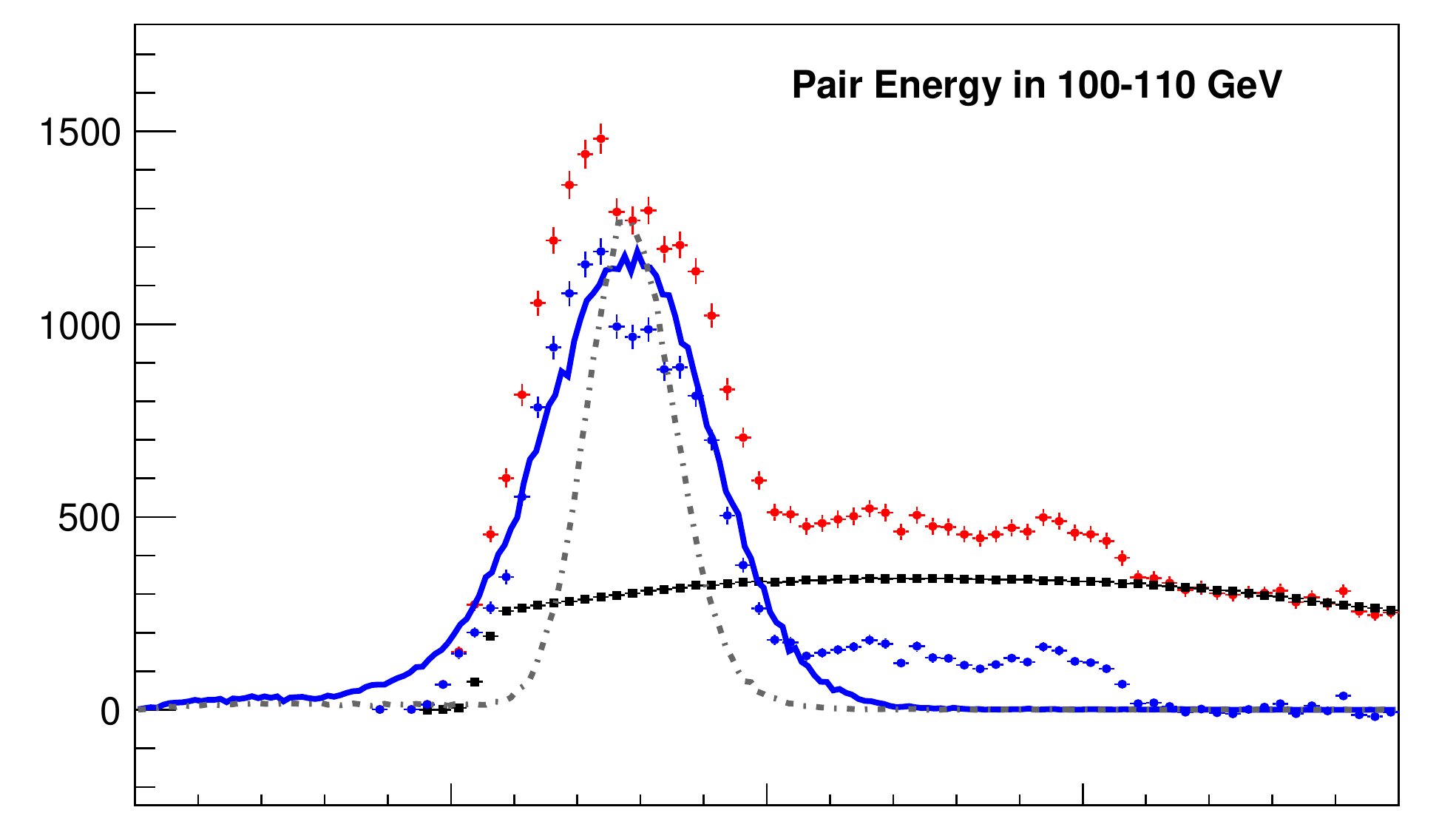}
 
 \includegraphics[width=61mm,clip]{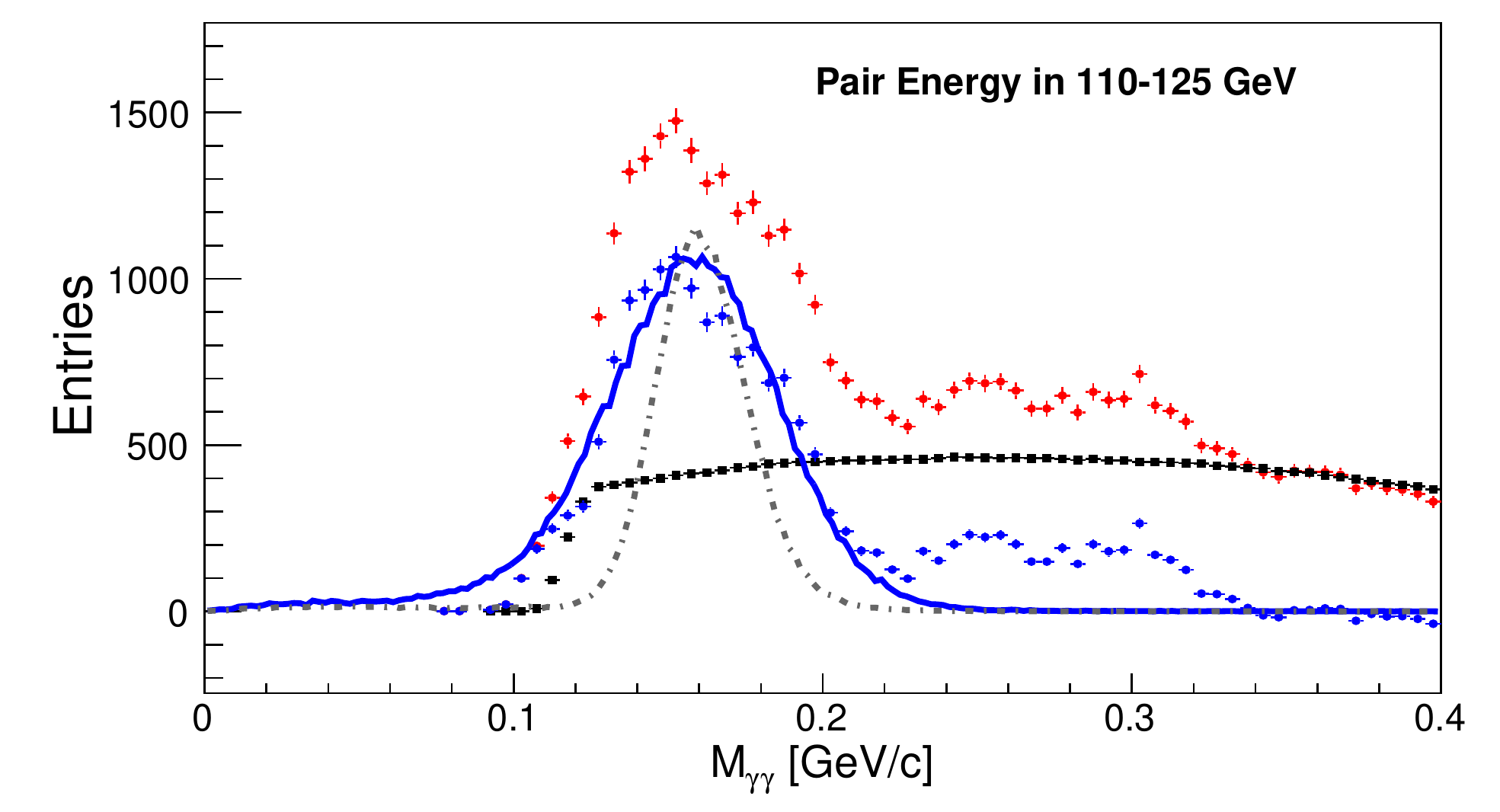}
 \includegraphics[width=58mm,clip]{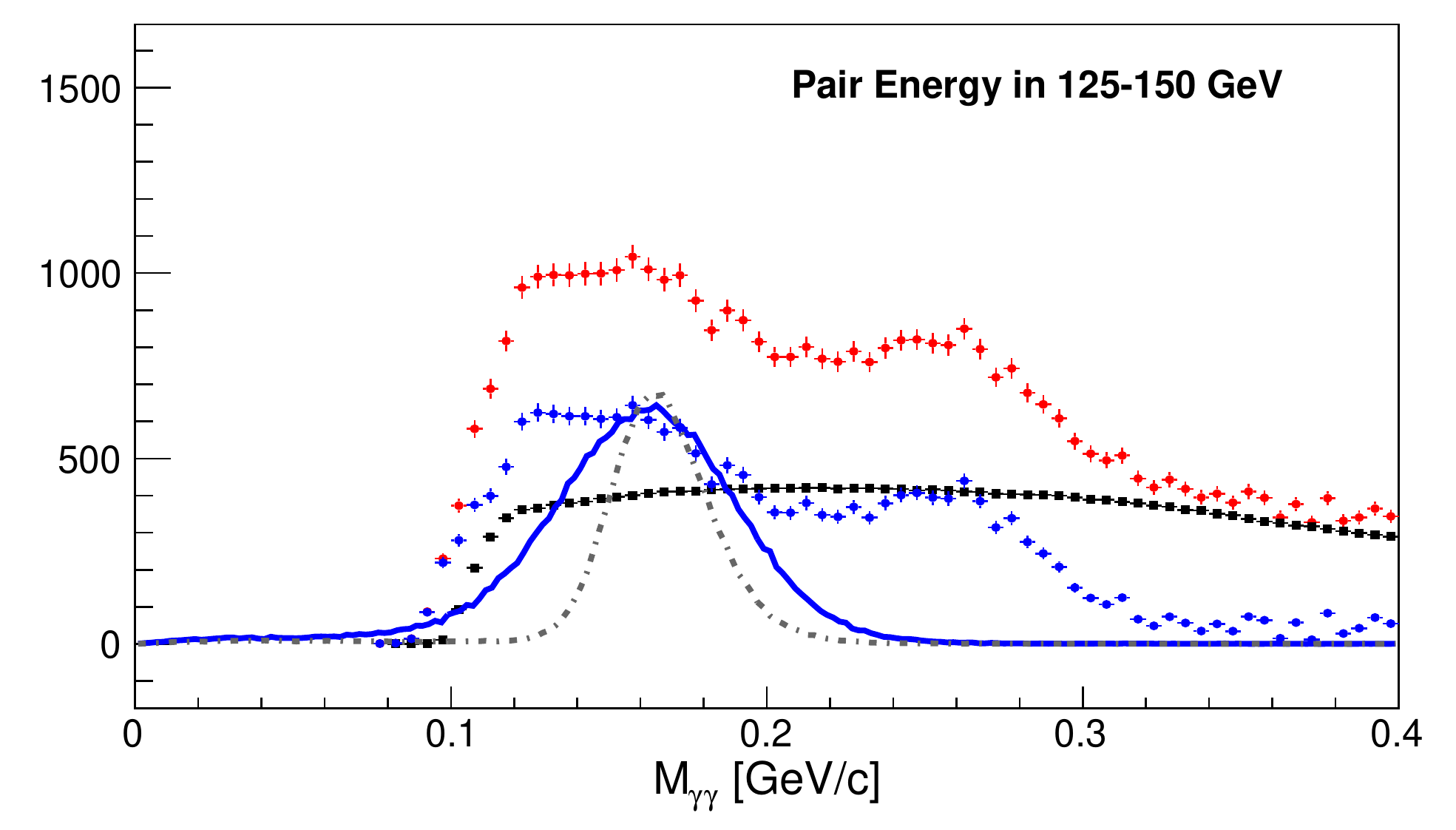}
\vspace{-5mm}
\caption{Invariant mass distributions of reconstructed cluster pairs in proton--proton collisions at $\sqrt{s}=$ 13 TeV in the Mini FoCal detector in different pair energy bins. 
Data are shown before and after a subtraction of the mixed-event background. 
The simulation results were reconstructed with realistic noise levels and the 12 layer readout as experienced in Mini FoCal, while also comparing to the full 20-layer setup with the same noise levels.}
\label{fig:InvMass}
\end{figure}

In addition to the single cluster reconstruction, we also attempted to reconstruct neutral pions via the di-cluster invariant mass calculation: $M_{\gamma\gamma} = \sqrt{2E_1E2(1-\cos{(\Theta_L)})}$, where $E_1$ and $E_2$ are the two cluster energies and $\Theta_L$ is the opening angle between the two clusters, assuming that the photons have originated at the ALICE interaction point. 
In order to reconstruct the central position of the cluster, we use a simple weighted mean of hits around the maximum: $\left<x\right> = \sum_i{x_i E_i}/\sum_i{E_i}$. 
Because of the limited two-shower separation capabilities of this pad-only prototype we require that the two clusters are at least $2.5~{\rm cm}$ apart from each other. 
The distributions of the reconstructed invariant mass in different pair energy bins is shown in \Fig{fig:InvMass}.

We developed a standalone~(i.e.~independent of the remaining ALICE simulation framework) Geant4 simulation to simulate single neutral pions generated 7.5 meters in front of our detector. 
Similarly as before, we implemented realistic noise in the simulation using only the first 12 layers as in the data. 
\Fig{fig:InvMass} also shows the distributions obtained from the simulation, for the realistic case using 12 layers~(blue) as well as for the nominal case using all 20 layers~(dashed gray).
The reconstructed mass distributions in data additionally contain combinatorial background from the physics collisions, which are not there in the simulation. 
This background was estimated using the mixed-event technique.
After subtraction of the mixed-event background, the mass peaks corresponding to the pions are well described by the realistic simulation. 
With 20 layers in the realistic simulation, which represent the case that the readout would have worked without communication errors, the invariant mass peak would be have significantly narrower as shown in \Fig{fig:InvMass}, most likely due to the better energy containment of the showers. 

However, for higher energies an additional structure appears in the measured distributions at higher masses, approximately at 0.25~GeV$/c^2$, which is not seen in the simulation. 
This might be a result of the additional material in the front of the detector, which is not fully modeled in the simulation, as e.g.\ the valves present on the beam pipe, the radiation monitor equipment, and related beam instrumentation. 
Additionally, there can be some in-jet correlation effecting the clusterization algorithm and creating an artificial peak. 
The efficiency to reconstruct pairs for higher energies is rapidly decreasing above 125~GeV, therefore we cannot observe any significant neutral pion peak above these energies.

In the final FoCal detector, two pixel layers will be installed with the main purpose to improve two-shower separation and thus the neutral pion reconstruction at high energy. 
The superior position resolution of those layers will also improve the mass resolution compared to a measurement for pads only, which is shown here.

\section{Summary}
\label{sec:summary}
We constructed a large-scale prototype of the electromagnetic calorimeter as a part of the FoCal project for the ALICE experiment at LHC. 
The Si-W prototype, which was built based on n-type sensors, was tested at the SPS and at LHC point 2 in ALICE. 
The energy resolution of the prototype is about 4.3\% for electron beams at both 150 and 250 GeV, consistent with results from realistic detector-response simulations obtained with Geant4.
Shower profiles in the longitudinal direction were also measured and found to be similar to those obtained from realistic simulations. 
The same prototype was also installed in the ALICE experimental area and and used to measure inclusive  cluster-energy distributions of electromagnetic showers in pseudorapidity $\eta$=3.7$--$4.5 at 7.5~m distance from the interaction point in proton--proton collisions at $\sqrt{s}$ = 13 TeV at LHC. 
The measured shower cluster energy distributions in different $\eta$ windows and the reconstructed di-cluster distributions, which dominantly originate from decay photons from neutral pions, were compared with those obtained in PYTHIA8 simulations.  
We found a reasonable agreement between data and simulation.


The construction of the Mini FoCal and its successful operation at about 7.5m of the ALICE interaction point in pp collisions at 13 TeV demonstrates the feasibility of the operation of the final FoCal under real-life conditions and with the large in-situ backgrounds in high-energy pp collisions at the LHC.

\section{Acknowledgments}
\label{sec:acknowledgments}
We thank H.~Muller and the RD51 collaboration at CERN for their kind support on the  readout and frontend electronics system using APV25 hybrid boards and SRS. 
We would also like to thank the staff members of the CERN accelerator complex for providing stable beams at PS and SPS beam tests. 
We thank the staff members of ELPH at Tohoku University for providing stable beams and kind support for this work. 
We thank the ALICE-DAQ team for helping with the DATE acquisition system; and the ALICE collaboration for the permission and support to place the prototype detector in the ALICE cavern. This work was supported by JSPS KAKENHI Grant Numbers JP17H01122, JP20H05638, JP26220707, JP19H01928.

\bibliographystyle{elsarticle-num}
\bibliography{references.bib}
\end{document}